\documentclass[reprint,amsmath,amssymb,pra]{revtex4-1}
\usepackage{times}
\usepackage{units}
\usepackage{ulem}
\usepackage{amsmath}
\usepackage{amssymb}
\usepackage{amsfonts}
\usepackage{epstopdf}
\usepackage{microtype}

\usepackage{graphicx} 			%include math packages
\usepackage{bm}				%include bold math package
\usepackage{color}			%include colored text support
\usepackage{pstricks}			%include postscript drawings support
\usepackage{float}			%include support for 'floating' objects (tables, figures, etc.)
\usepackage{hyperref}
\newcommand{\Div}[1]{
\vec{\nabla} \cdot  \vec{#1} 
}
\newcommand{\Grad}{
\vec{\nabla} 
}

\newcommand{\ltm}{LTM}
\newcommand{\ltms}{LTMs}

\begin{document}

\title{Non-Hermitian coupled-mode theory for incoherently pumped exciton-polariton condensates}
\author{S. Khan}
\author{H. E. T\"ureci}
%\author{\textsf{author3}}
\affiliation{Department of Electrical Engineering, Princeton University, Princeton, New Jersey 08544, USA}
\date{\today}

\begin{abstract}
The generalized Gross-Pitaevskii equation (gGPE) is an effective phenomenological description for the dynamics of incoherently pumped exciton-polariton condensates. However, a brute force numerical simulation of the gGPE provides little physical insight into condensate formation under arbitrary pumping configurations, and is demanding in terms of computational resources. We introduce in this paper a modal description of polariton condensation under incoherent pumping of arbitrary spatial profile, based on eigenmodes of the non-Hermitian generator of the linearized dynamics. A pump-dependent basis is then introduced to formulate a temporal coupled-mode theory that captures condensate dynamics in the presence of all nonlinear interactions. Simulations using a single set of modes for a given pumping and trapping configuration agree very well with a full integration of the gGPE in diverse dynamical regimes, supporting the validity of this modal description, while also providing a speedup in simulation times.
\end{abstract}

\maketitle

\section{Introduction}

Exciton-polaritons, quasiparticles that arise from the strong coupling of cavity-confined photons and excitons confined in quantum wells, are effectively bosonic at low densities and can undergo condensation under incoherent excitation beyond a critical pump power~\cite{kasprzak_boseeinstein_2006, balili_bose-einstein_2007, deng_polariton_2003}. A theoretical description of the spatio-temporal dynamics of the associated condensate order parameter can be obtained through a generalized Gross-Pitaevskii equation (gGPE) introduced in Refs.~\cite{wouters_excitations_2007, keeling_spontaneous_2008}. While the microscopic basis of such a description starting from a quantum field theory still remains at large~\cite{carusotto_quantum_2013}, as a phenomenological model its strength in describing many recent experiments under various pumping and trapping conditions has proven to be quite remarkable. Despite this success however, this approach remains a computational simulation tool that is often invoked a posteriori to justify experimentally observed condensation dynamics~\cite{lagoudakis_quantized_2008, amo_collective_2009, amo_superfluidity_2009, lagoudakis_observation_2009, sanvitto_persistent_2010, manni_spontaneous_2011, nardin_hydrodynamic_2011, sanvitto_all-optical_2011, askitopoulos_polariton_2013, nguyen_acoustic_2015}. Its limitation in providing physical insight and predictive power is arguably due to a lack of an associated modal description. In lasers, the projection of the laser field on resonator modes generally provides valuable insight into resulting spatial patterns of lasing modes in their steady-state, their thresholds and frequencies. In addition, a modal projection is typically computationally very efficient, turning a coupled set of non-linear partial differential equations into a set of ordinary differential equations for which there are many standard and powerful algorithms. 

\begin{figure}[t]
\includegraphics[scale=0.252]{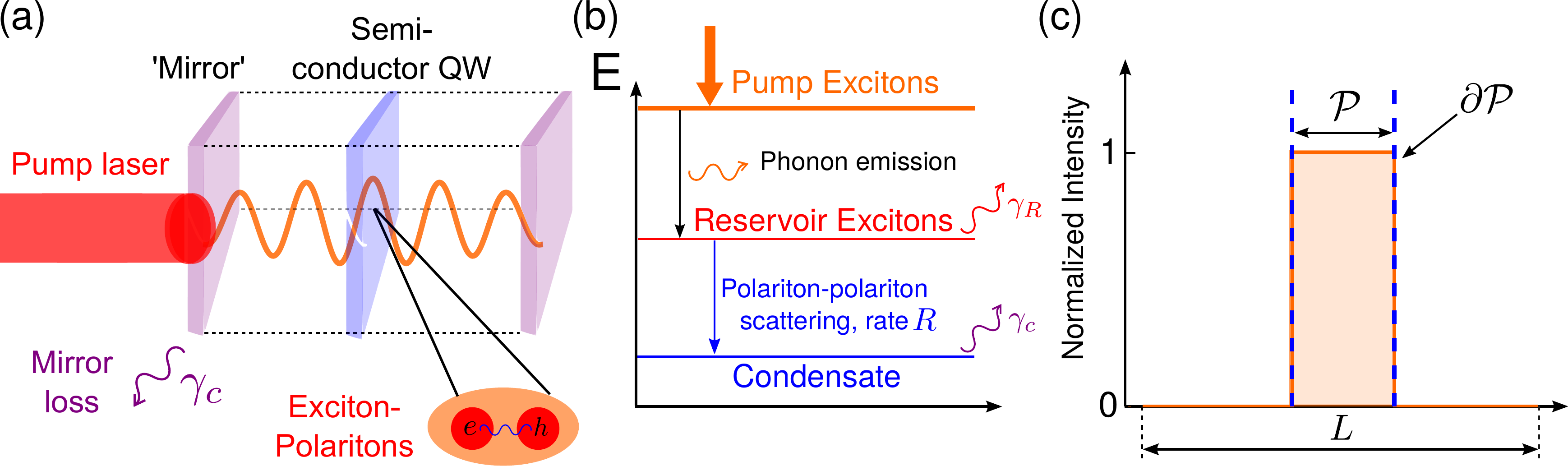}
\caption{ (a) Schematic diagram of pumped microcavity with embedded semiconductor quantum wells, under incoherent pumping. (b) Schematic representation of the mechanism that populates the exciton reservoir under incoherent pumping. (c) Pump profile and normalized intensity for uniform pump in 1D. Dashed lines indicate the pump boundary $\partial\mathcal{P}$, beyond which the pump intensity per unit area, $Pf(\mathbf{r})$, is zero. The region enclosed by this boundary is defined as the pump region $\mathcal{P}$.}
\label{f:pumpRegion}
\end{figure}

For trapped atomic gas systems, condensation typically occurs in the ground state of the trapping potential. When studying dynamics of trapped gases, the normal modes of the trapping potential provide a suitable basis to project the gGPE onto, resulting in a set of coupled (nonlinear) ordinary differential equations for the dynamical projection coefficients. The Hermitian nature of the Hamiltonian ensures that a complete and orthogonal basis exists, while the energy eigenvalues of the modes provide an organizing principle that allows the proper truncation of the basis in the infinite dimensional Hilbert space: one orders the modes according to their energy and includes only modes with energies less than a suitably chosen energy cutoff. One source of the difficulty in arriving at a modal description appropriate for the non-equilibrium condensate dynamics is that many recent experiments employ large-area cavities (with the exception of `0D polariton boxes'~\cite{balili_bose-einstein_2007, ferrier_polariton_2010, nardin_phase-resolved_2010, ferrier_interactions_2011, kalevich_ring-shaped_2014, besga_polariton_2015, zhang_coupling_2015, dufferwiel_tunable_2015}) where the trapping of the condensate in the plane is achieved via the finite extent of the pump beam (see Fig.~\ref{f:pumpRegion}~(a)). In this setting the expansion of the polariton field in resonator modes indexed by a discrete set of quasi-momenta is not very useful for two reasons. First, the energy eigenvalues found through this procedure do not provide a good truncation scheme because the condensation can typically occur into a higher energy (= higher momentum) state. Second, there is a net energy flux through the system, from the pump to the reservoirs and a Hermitian description is out of the question or can at best be approximate under ideal conditions (of very high Q, spatially highly extended uniform pump and a very large, translationally symmetric structure in the plane). 

Because of the above-stated conditions, in condensates that are trapped optically (i.e. by a spatially patterned incoherent pump beam~\cite{tosi_sculpting_2012, manni_spontaneous_2011, askitopoulos_polariton_2013, gao_observation_2015, baboux_bosonic_2016, sun_stable_2016}), it is often not clear what the proper ``energy levels" are that the excitations generated by the pump may condense into. From the more traditional point of view looking at the condensation in terms of momentum modes, motivated by the approximate translational invariance of the sample in the absence of the pump spot, condensation into a complex spatial pattern may appear as condensation into a linear superposition of momentum eigenstates. This often can lead to the misconception that there is something more profound taking place then condensation into a pure momentum mode, such as synchronization. 

Clearly, the appropriate set of modes have to be closely related to the geometry of the pump beam. However, the reservoir polaritons generated by the pump act not only as an energetic barrier but also as a source of gain for the condensing polaritons. One is then led to conclude that the appropriate set of modes are subject to a complex-valued potential that is pump-dependent. An approach to the description of the condensate dynamics in terms of the modes of a non-Hermitian operator was introduced in Ref.~\cite{ge_pattern_2013}, and the insight and predictive power it provides was demonstrated in recent experiments~\cite{sun_stable_2016, baboux_bosonic_2016}. The method presented in Ref.~\cite{ge_pattern_2013} is limited to capturing the steady-state dynamics of the gGPE (when such exists). Many interesting dynamical phenomena, such as condensate synchronization~\cite{baas_synchronized_2008, wouters_synchronized_2008, eastham_mode_2008} and self-pulsing, or transient dynamics, remain beyond the reach of this approach. The goal of the present work is to discuss a non-Hermitian coupled mode theory to address this shortcoming. 

While a complete non-Hermitian basis can be consistently defined to project the gGPE on, a more subtle question concerns the associated organizing principle that will provide the most effective truncation scheme. There is no unique way of doing this, but we show here that there is a set of \textit{pump modes} associated with the {\it unsaturated pump (reservoir) distribution} in the absence of polariton interactions, which provides maximal physical insight. Pump modes are a set of non-Hermitian modes that parametrically evolve with the pump strength parameter $P$. This is physically sensible because the stronger the pump the higher are the barriers of the potential generated by the deposited exciton reservoir (which we assume to be immobile). We show that the members of this set are analytically connected to the exact condensate modes in the absence of nonlinear interactions and gain depletion; we refer to the latter as linear threshold modes (\ltms{}). \ltms{} therefore provide the organizing principle that is required to effectively truncate the pump modes. Eigenvalues of the associated non-Hermitian dynamical evolution operator provide insight into the cumulative gain the associated patterns experience and their associated condensation energies (frequencies) and thresholds. 

The condensation process appears then as an instability of the fluctuation patterns described by the pump modes. In fact, the first member of the set of pump modes provides the {\it exact} threshold, frequency and spatial distribution of the first condensing mode. We show subsequently that this approach can be extended far into the strong-pumping regime where nonlinear interactions dominate the dynamics. The pump modes provide a suitable basis for an efficient projection of the full condensate wavefunction, with time-dependent coefficients, forming the foundation for the nonlinear temporal coupled mode theory (TCMT) that we introduce in this paper. The TCMT can accurately capture condensate and reservoir dynamics in the presence of all nonlinear effects and interactions, whether a steady-state exists or not. By decoupling the computation of the spatial modes from the temporal evolution of the system, the operative equations of the TCMT are reduced to a set of coupled ordinary differential equations (ODEs). This provides computational advantages, and should be particularly advantageous for geometries in more than one dimension. Most importantly, simulation results using the TCMT agree very well with a full integration of the gGPE using a split-step symplectic integrator (SSI). This places our modal theory on sound footing as an appropriate description of nonlinear condensate dynamics under incoherent pumping. 

The rest of this paper is organized as follows: in Sec.~\ref{sec:linear}, after a brief review of the standard mean-field description of incoherently pumped polariton condensates, we introduce the linear threshold modes (\ltms{}) to discuss condensation in an arbitrary pump-field and resonator geometry. In Sec.~\ref{sec:tcmt} we use this modal description as an organization principle to develop the pump basis, a non-Hermitian, pump-power dependent basis; this functions as the basic building block of our modal theory. Employing this basis, we derive a set of dynamical equations that constitute the TCMT. Finally, Sec.~\ref{sec:numerics} is reserved for numerical comparisons of the TCMT with the SSI, for an all-optically trapped condensate, as well as condensation in pumped microcavity systems, to highlight agreement across various dynamical regimes.

\section{Linear Threshold Modes}
\label{sec:linear}

In this section, we discuss the first step towards a full modal description for the gGPE through the pump basis. The linear threshold modes (\ltms{}) introduced in Sec.~\ref{sec:LTM} will provide a scheme to index the pump modes that constitute this basis; this discussion can be found in Sec.~\ref{sec:basis}. Before introducing the \ltms{},  we present a brief review of the standard mean-field description of incoherently pumped polariton condensates \cite{deng_exciton-polariton_2010, carusotto_quantum_2013} to establish our notation and lay the foundations for the analysis to follow.  

\subsection{Mean-field description of incoherently pumped polaritons}

Inorganic exciton-polaritons (referred to as polaritons from this point on) are quasiparticles of cavity photons and semiconductor excitons generated for example when a semiconductor quantum well is placed within a light-confining microcavity (see Fig.~\ref{f:pumpRegion}~(a)). Polaritons with inorganic excitonic components exhibit strong Coulomb-like repulsions with each other; as such, these polaritons behave essentially like photons dressed with a very light mass (typically 10\textsuperscript{-4}$m_e$) and interactions. The strong interactions make for a nontrivial many-body problem; the standard first method of analysis is therefore a mean-field treatment of the interactions based on the Gross-Pitaevksii equation~\cite{dalfovo_theory_1999}, first developed for the study of condensates of trapped gases.

Different from the case of trapped gases is the need to account for polariton losses and hence for continuous pumping to overcome these losses. The pump generates a reservoir of high energy excitons with population density $n_R(\mathbf{r},t)$ that scatter continuously into lower energy polaritons; this amplification must overcome losses for a condensate phase to be possible (see Fig.~\ref{f:pumpRegion}~(b)). Then, the generalized Gross-Pitaevskii equation (gGPE) for the condensate wavefunction $\Psi(\mathbf{r},t)$~\cite{wouters_excitations_2007, deng_exciton-polariton_2010, carusotto_quantum_2013} is:
\begin{align}
i\frac{\partial }{\partial t}\Psi = \left[ -\frac{\nabla^2}{2m} + g_R n_R + g|\Psi|^2 + \frac{i}{2}\left(Rn_R - \gamma_c\right) \right] \Psi,
\label{NGPE}
\end{align}
which takes the form of a nonlinear Schr\"odinger equation (we set $\hbar=1$ throughout this paper). Polariton-polariton repulsions $\propto g$) and repulsion between condensate polaritons and reservoir excitons ($\propto g_R$) are described as contact interactions, and are treated at the mean-field level. Polariton systems experience continuous losses, as the photonic fraction of polaritons leads to leakage from the confining cavity. This openness is encapsulated by the non-Hermitian parts of effective Hamiltonian describing condensate evolution. Here, $\gamma_c$ is the out-of-plane (or mirror) loss and $R$ the amplification rate for stimulated scattering from the reservoir. Scattering of polaritons from this reservoir into the condensate~\cite{porras_polariton_2002} overcomes losses beyond a critical pump power, and a condensate forms (see Fig.~\ref{f:pumpRegion}~(b) for a schematic).

The exciton reservoir also evolves dynamically as the condensate forms, and therefore the reservoir density $n_R(\mathbf{r},t)$ has its own governing equation~\cite{wouters_excitations_2007}:
\begin{align}
\frac{\partial}{\partial t}n_R = Pf(\mathbf{r}) - \gamma_R n_R - Rn_R|\Psi|^2,
\label{res}
\end{align}
The first term accounts for the generation of the exciton reservoir by an external pump. We define a pump region $\mathcal{P}$ as being enclosed by the minimal boundary $\partial\mathcal{P}$ beyond which the pump intensity vanishes; see Fig.~\ref{f:pumpRegion}~(c) for an example in a 1-dimensional system. The spatial pump profile $f(\mathbf{r})$ is unit normalized over this pump region, so that the pump strength parameter $P$ is the total number of excitons generated within $\mathcal{P}$ per unit time. The term proportional to the scattering rate $R$ accounts for the depletion of the reservoir population as polaritons scatter into the condensate. Any losses from mechanisms other than scattering into the condensate (e.g. exciton recombination losses) are given by $\gamma_R$.

The full dynamics of incoherently pumped polaritons are therefore described by the set of coupled nonlinear partial differential equations (PDEs), Eqs.~(\ref{NGPE}),~(\ref{res}). Solving this system requires an integration over a spatio-temporal grid. The standard approach is to employ a symplectic split-step integrator (SSI) based on separate real and momentum space evolution, and while this technique has proven useful, it has some significant shortcomings. Firstly, and most importantly, a full integration of the coupled PDEs using a SSI provides little general information about spatio-temporal condensate dynamics beyond the results that are directly simulated. The simulations also bear computational difficulties: spatio-temporal integration of PDEs is much more resource-consuming than the integration of ordinary differential equations (ODEs), while the use of Fast Fourier transforms (FFTs) in the SSI algorithm forces spatial periodicity onto a system where there may be none. This can introduce computational artifacts unless a large enough spatial domain is used, which incurs significant computational slowdown and/or limited spatial resolution. 

With the aim of addressing these challenges, the next sections will introduce an associated modal description of polariton condensation under incoherent pumping. Such a description provides more insight into condensate formation and dynamics under general pumping and trapping configurations, and will also be shown to hold in the nonlinear regime.

\subsection{Defining the Linear Threshold Modes}
\label{sec:LTM}

In studying the condensate dynamics at a pump power $P$ as described by Eqs.~(\ref{NGPE}),~(\ref{res}), it is reasonable to first consider the simplified problem of an unsaturated reservoir in the absence of polariton interactions i.e. we drop nonlinear terms $\propto |\Psi|^2$ in Eqs.~(\ref{NGPE}),~(\ref{res}). This yields the linear problem:
\begin{align}
i\frac{\partial}{\partial t}\Psi = \left[ -\frac{\nabla^2}{2m} + s Pf(\mathbf{r}) - \frac{i}{2}\gamma_c \right]\Psi \equiv \mathcal{H}_{\rm L}(P)\Psi 
\label{linGPE}
\end{align}
The dynamics of excitations under these conditions are entirely determined by the linear, non-Hermitian operator $\mathcal{H}_{\rm L}(P)$. We have introduced the pump-induced potential $s$,
\begin{align}
s = \frac{1}{\gamma_R}\left(g_R + \frac{i}{2}R \right)
\label{s}
\end{align}
which encapsulates the effect of the pump-generated reservoir in this linear regime: a repulsive potential and a source of gain, with a spatial dependence given by the pump profile. We now consider fluctuations $\varphi(\mathbf{r})$ of this operator,
\begin{align}
\Psi(\mathbf{r},t) = \varphi(\mathbf{r})e^{-i\nu t}
\label{fluctuation}
\end{align}
Substituting this ansatz into Eq.~(\ref{linGPE}) immediately gives:
\begin{align}
\mathcal{H}_{\rm L}(P) \varphi_n(\mathbf{r};P) = \nu_n(P) \varphi_n(\mathbf{r};P)
\label{HLP}
\end{align}
indicating that $\varphi$ is an eigenmode of $\mathcal{H}_{\rm L}(P)$, with eigenvalue~$\nu$. In the general case, we choose the pump boundary $\partial\mathcal{P}$ as circumscribing the minimal region that encompasses all pump regions and trapping potentials; then, the polariton field beyond $\partial\mathcal{P}$ will be outgoing only, carrying an outward flux of polaritons that decays with distance from the pump region (due to nonzero out-of-plane loss $\gamma_c$). While a boundary beyond $\partial\mathcal{P}$ may also be chosen, the minimal choice is computationally most efficient. The outgoing flux solution must be continuously connected to the polariton field inside the pump region via a {\it boundary condition} at $\partial\mathcal{P}$. For concreteness, we will discuss here the 1D case for which this condition can be written as
\begin{align}
\partial_x \varphi_n |_{\partial\mathcal{P}^\pm} = \pm i q(\nu_n) \varphi |_{\partial\mathcal{P}^\pm}
\label{bc}
\end{align}
where $q^2(\nu_n)/2m = \nu_n + i\frac{\gamma_c}{2}$ is the outgoing (complex) wavevector, and $\pm$ denotes the condition for the right and left pump boundary respectively. An analogous condition can be found in 2D as well. Restricting the computational domain to within $\partial\mathcal{P}$ requires solving a non-Hermitian boundary value problem that parametrically depends on $P$. The eigenvalues $\nu_n$ are discrete and generally complex-valued. According to our definition in Eq.~(\ref{fluctuation}), decaying solutions for such fluctuations are confined to the lower complex plane; thus, the positive/ and negative imaginary parts of the eigenvalues represent the net gain and loss respectively experienced by the corresponding fluctuation with the spatial pattern $\varphi_n$, while the real parts correspond to the associated mode frequencies. The physics of the linear problem is built into these eigenvalues: generally, as a function of an increasing value of $P$, the eigenvalues move toward higher frequencies due to the increasing pump-induced blueshift, while the imaginary parts of all eigenvalues become less negative as the gain increases, flowing towards the real line. 

For a given $P=P_n$, one of the eigenvalues intersects the real line, i.e. $\nu_n$ is real (the rest of the eigenvalues at this power are generally complex-valued); for a pump power infinitesimally beyond $P_n$, the corresponding fluctuation $\varphi_n(\mathbf{r};P_n)$ becomes unstable, which describes the onset of condensation in this formulation. It is possible to show that there are $N$ pairs $(\omega_n = \nu_n(P_n), P_n)$ for which the $\omega_n$ are real-valued. Here $N$ is the dimension of the linear problem for which we will present a systematic truncation scheme in later sections. We refer to these modes as \textit{linear threshold modes} (\ltms{}) and index them in the order of increasing $P_n$, also introducing a specific notation: $\varphi_n^{\rm L} \equiv \varphi_n^{\rm L}(\mathbf{r};P_n^{\rm L},\omega_n^{\rm L})$ is referred to as the $n^{\rm th}$ linear threshold mode (\ltm{}), with linear threshold power $P_n^{\rm L}$ and real frequency $\omega_n^{\rm L}$. The $n^{\rm th}$ \ltm{} is therefore a special eigenmode of $\mathcal{H}_L(P_n^{\rm L})$, with a purely real eigenvalue (on account of it being at threshold) representing the condensation frequency. 

\subsection{Significance of Linear Threshold Modes}

Our discussion of \ltms{} above answers the question of how to define condensate modes under incoherent pumping when the confining geometry is determined entirely by the pump. Furthermore, the \ltm{} with lowest power threshold has special meaning: it corresponds to the first unstable fluctuation of the uncondensed $\Psi = 0$ phase as the pump power is increased. As such, it provides {\it exactly} the field distribution and the frequency of the mode that is observed when condensation first occurs. Beyond the corresponding threshold pump power, $P_1^{\rm L}$, nonlinear effects become important and the linearized theory is no longer valid.  

As such, the significance of \ltms{} beyond the lowest power threshold \ltm{} may not be immediately obvious, since these are strictly speaking objects relevant to the linearized theory. However, it can be shown that the linearized power thresholds organize the \ltms{} into a hierarchy, which is directly related to how efficiently each pump-confined mode utilizes the pump for gain~\cite{ge_pattern_2013, sun_stable_2016}. The $n=1$ \ltm{} has lowest power threshold and is most efficient at utilizing the pump, the $n=2$ mode is second most efficient, and so on. Therefore, the \ltms{} are in a quantifiable way the preferred spatial configurations of polaritons under a given pumping configuration. We will see next that it is possible to define a non-Hermitian pump-dependent basis that can be used to efficiently project gGPE on. The \ltm{}s are then used to provide an efficient truncation scheme for this basis. 

\section{Nonlinear Temporal Coupled Mode Theory}
\label{sec:tcmt}

In this section, we develop the central result of this paper: a temporal coupled-mode theory (TCMT) for the fully nonlinear, time-dependent problem of polariton condensation under incoherent pumping. The first task is to determine an appropriate basis for expansion of the nonlinear condensate wavefunction; we will find that specific eigenmodes of $\mathcal{H}_{\rm L}(P)$ that can be continuously connected to the \ltms{} provide such a basis; we refer to these as pump modes and introduce them next. Then, by making an expansion of the condensate wavefunction in this basis, we derive the dynamical coupled-mode equations, presented in Eqs.~(\ref{TCMTa}),~(\ref{TCMTb}).

\subsection{The Pump Basis}
\label{sec:basis}

To derive a coupled mode theory for Eqs.~(\ref{NGPE}),~(\ref{res}) that can capture general time-dependent behavior, we must expand the condensate wavefunction in a complete set of spatial modes with time-dependent coefficients. Intuitively, an expansion in eigenmodes of $\mathcal{H}_{\rm L}(P)$ makes sense, since these modes already take into account the complexity of the linearized condensation problem for arbitrary pump profiles. However a consistent definition of this basis requires the generalization of the problem to a two-parameter family of boundary value problems (BVPs), for which we use the notation $\mathcal{H}_{\rm L}(P, \Omega)$. This BVP is defined by the same differential operator in Eq.~\ref{HLP}, but with the parametric boundary condition 
\begin{align}
\partial_x \varphi_n |_{\partial\mathcal{P}^\pm} = \pm i q(\Omega) \varphi |_{\partial\mathcal{P}^\pm}
\label{bcPB}
\end{align}
The solutions of this BVP describe self-sustained oscillations of the condensate at (real) frequency $\Omega$. Hence, the set of $\varphi_n$'s parametrically depends on two real parameters $(P,\Omega)$. Being eigenmodes of the same non-Hermitian operator with the same boundary condition, the modes $\{\varphi_n(\mathbf{r};P,\Omega)\}$ can be shown to form a complete, biorthogonal basis, satisfying an orthogonality relation with an unconjugated inner product~\cite{tureci_ab_2009}:
\begin{align}
\int_{\mathcal{P} } d\mathbf{r}~\varphi_n(\mathbf{r};P,\Omega)\varphi_m(\mathbf{r};P,\Omega) = \delta_{nm}
\label{orth}
\end{align}
Note that the orthogonality relation differs from the usual power orthogonality involving a conjugated inner product due to the non-Hermiticity of the modes. 

Next we use this basis to expand the full condensate wavefunction in Eq.~(\ref{NGPE}):
\begin{align}
\Psi(\mathbf{r},t) = \sum_{n=1}^N a_n(t) \varphi_n(\mathbf{r}; P,\Omega)e^{-i\Omega t}
\label{exp}
\end{align}
where $a_n(t)$ are time-dependent basis coefficients. Here, the purpose of $\Omega$ is to provide an optimal rotating frame in which to express the basis coefficients $a_n(t)$. While there is no unique way to choose this frequency, some choices yield better results than others. In due course, we will present a suitable optimization procedure. 

We now address the more subtle question of an effective and physically transparent truncation scheme for the expansion in Eq.~(\ref{exp}). For this we first make the following observation: when $P=P_n^{\rm L}$ and $\Omega=\omega_n^{\rm L}$, one member of the basis set exactly coincides with the $n^{\rm th}$ \ltm{}, $\varphi_n^{\rm L}(\mathbf{r};P_n^{\rm L},\omega_n^{\rm L})$. It can further be shown that the eigenvalue flow as a function of $(P,\Omega)$ is generally differentiable so that for a general $(P,\Omega)$ one of the modes can be continuously connected to this particular \ltm{}. This provides then a very effective principle to organize the modes (i.e. index them): the $n^{\rm th}$ basis mode is connected to the $n^{\rm th}$ \ltm{}. Recall that this \ltm{} is the $n^{\rm th}$ most effective at utilizing the pump for gain. Thus, a truncation performed at a judiciously chosen upper cutoff will leave an ordered basis of the first $N$ optimal modes that experience gain from the deposited exciton reservoir defined by $f(\mathbf{r})$. We call this so-truncated, ordered basis the \textit{pump basis}. The $N$ \textit{pump modes} that constitute this ordered basis are simply eigenmodes of $\mathcal{H}_{\rm L}(P,\Omega)$ chosen such that they connect to the first $N$ \ltms{}. 

\begin{figure}[t]
\includegraphics[scale=0.23]{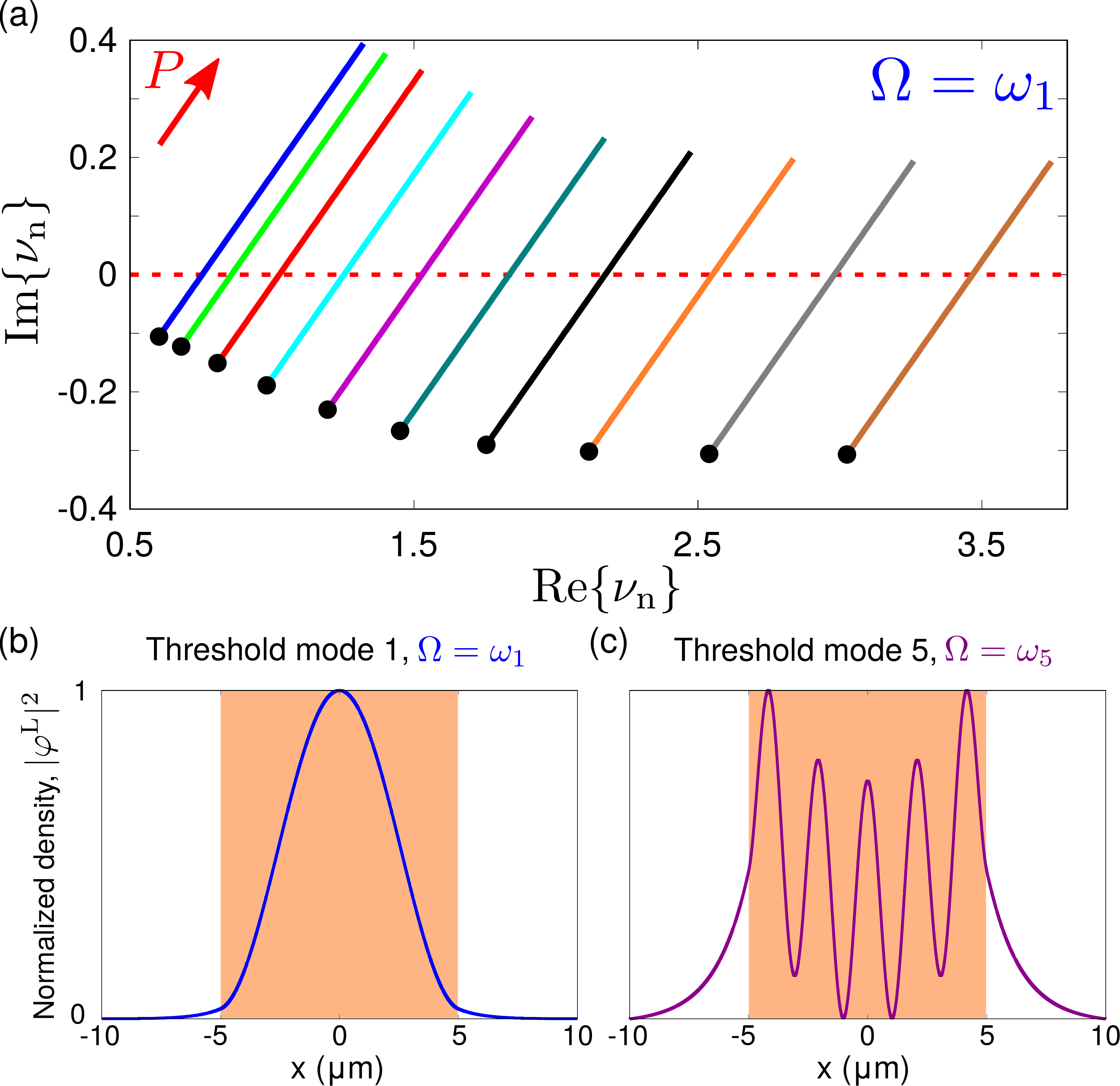}
\caption{(a) Trajectory of eigenvalues $\nu_n(P),~n\in[1,10]$ in the complex plane as pump power is increased, for fixed outgoing frequency $\Omega = \omega_1$. The leftmost trajectory is for $n=1$, the rightmost for $n=10$, and the arrow indicates the direction of flow. The real axis projection of the $n$th trajectory is the real frequency, while the imaginary axis projection is the loss or gain. (b) Plots of the \ltms{} for $n=1$ and (c) $n=5$ respectively. The orange shaded region indicates the uniform pump.}
\label{f:eigenComplex}
\end{figure}

To be able to make the connection between the pump modes and \ltms{}, we need an efficient computational procedure for the \ltms{}, which we present now. For concreteness, we discuss the case of a 1D uniform pump spot~(see Fig.~\ref{f:pumpRegion}~(c)). We first fix $\Omega$ that defines the boundary condition of our BVP through Eq.~(\ref{bcPB}). This is our best guess for one of the \ltm{} frequencies. Some typical eigenvalue trajectories with increasing pump power $P$ are shown in the complex-$\nu$ plane in Fig.~\ref{f:eigenComplex}~(a). For a low enough pump power, all calculated eigenmodes have eigenvalues with negative imaginary parts; this is the regime below condensation threshold, where any fluctuations around the $\Psi = 0$ state decay to zero for long times. As the pump power is increased, the net loss decreases for all modes, and the imaginary parts of all eigenvalues become less negative, flowing towards the real line, while the real parts move towards higher frequencies. As mentioned earlier, these features of eigenvalue flow are a manifestation of the pump-dependent physics in the linear regime, and hence holds for general pump profiles (although the specific trajectories may be more complicated). When the $n^{\rm th}$ mode reaches threshold and intersects the real line, it acquires a purely real eigenvalue $\omega_n$, representing its frequency. If $\Omega$ is different from $\omega_n$, the outlined procedure can be iteratively repeated until this constraint is satisfied (additional practical details of how to optimize this procedure can be found in Appendix~\ref{app:compModes}). The procedure for connecting the eigenmodes of $\mathcal{H}_{\rm L}(P)$ to the \ltms{} is then a simple matter of overlap minimization at a given $(P,\Omega)$, and is also described in Appendix~\ref{app:compModes}.

Lastly, we discuss the choice of $\Omega$ in Eq.~(\ref{exp}), which sets the boundary condition defining the pump modes via Eq.~(\ref{bcPB}). This has a simple interpretation: physically, $\Omega$ defines the condensate frequency. Eq.~(\ref{bcPB}) simply ensures that the pump modes obey the correct dispersion relation for an outgoing, decaying polariton flux emanating from a condensate with frequency $\Omega$. Crucially, this condensate frequency evolves with pump power due to the pump-induced potential and polariton-polariton interactions, and so $\Omega = \Omega(P)$. By imposing Eq.~(\ref{bcPB}), the pump modes take this evolution into account, a parameterization that proves crucial for accurate simulation results. We also explicitly extract time evolution at $\Omega$ in Eq.~(\ref{exp}) to render the $\{a_n(t)\}$ slowly-varying, which is efficient for numerical simulations. To determine $\Omega$, we employ a self-consistent procedure where we initially set $\Omega=\omega_1^{\rm L}$, namely the lowest threshold frequency determined before. The TCMT equations to be discussed further below are then solved. The most dominant frequency in the Fourier transform of the amplitudes in the long-time limit is then used to update $\Omega$. In practice, this procedure is found to have very good convergence properties.

The important outstanding question now is whether the pump basis is suitable for the analysis of the full nonlinear problem. To answer this, we will first derive the nonlinear coupled mode theory based on the expansion in Eq.~(\ref{exp}). Tests of this theory against a full simulation of Eqs.~(\ref{NGPE}),~(\ref{res}) will determine the validity of our formulation. 

\subsection{Coupled-Mode Equations}
\label{sec:TCMT}

The fundamental procedure for obtaining a coupled-mode theory from the gGPE-reservoir equations [Eqs.~(\ref{NGPE}),~(\ref{res})] is simple: we expand the condensate wavefunction $\Psi$ in the pump basis derived in the previous section (see Eq.~(\ref{exp})), and then integrate out the spatial dependence to obtain dynamical equations for the time-dependent expansion coefficients. This section implements this procedure, while clarifying additional subtleties that arise. However, before substituting the expansion for $\Psi$ in Eqs.~(\ref{NGPE}),~(\ref{res}), it is useful to rewrite these equations in a more convenient form. First, we extract explicitly the linear, time-independent part of the reservoir density:
\begin{align}
n_R(\mathbf{r},t) = \frac{Pf(\mathbf{r})}{\gamma_R}  + \widetilde{n}_R(\mathbf{r},t)
\label{displaceRes}
\end{align}
Our choice of `displacing' the reservoir density in this way - an {\it exact} transformation - allows the gGPE, Eq.~(\ref{NGPE}), to be written in a particularly transparent form:
\begin{align}
i\frac{\partial}{\partial t}\Psi = \mathcal{H}_{\rm L}(P)\Psi + \mathcal{V}_{\rm NL}(P) \Psi
\label{sepGPE}
\end{align}
where the linear and nonlinear terms in the condensate evolution are neatly separated; $\mathcal{H}_{\rm L}(P)$ is the linear generator introduced in Eq.~(\ref{linGPE}) which encapsulates the complex pump-induced potential and polariton loss, while the nonlinear potential $\mathcal{V}_{\rm NL}(P)$ is given by:
\begin{align}
\mathcal{V}_{\rm NL}(P) = \Bigg[ s\gamma_R \widetilde{n}_R(\mathbf{r},t) + g|\Psi|^2 \Bigg] \Psi
\label{VNL}
\end{align}
$\mathcal{V}_{\rm NL}(P)$ includes the real-valued polariton-polariton repulsion, but also the nonlinear, dynamical effects introduced by reservoir evolution, namely the saturating nonlinearity and reservoir depletion, via $\widetilde{n}_R(\mathbf{r},t)$. 

Now, using the expansion of the condensate wavefunction in Eq.~(\ref{exp}), the nonlinear gGPE can be cast as a set of coupled equations for the time-dependent coefficients $a_n(t)$. Since the pump modes that constitute our basis of expansion are eigenmodes of $\mathcal{H}_{\rm L}(P)$, the linear part of Eq.~(\ref{sepGPE}) simplifies greatly [c.f. Eq.~(\ref{HLP})]:
\begin{align}
\mathcal{H}_{\rm L}(P)\Psi = \sum_n a_n(t)~\nu_n(P) \varphi_n(\mathbf{r};P,\Omega)e^{-i\Omega t}
\label{HLPact}
\end{align}
The nonlinear potential $\mathcal{V}_{\rm NL}(P)$ can also be easily written in terms of the mode coefficients using the basis expansion. Equations of motion for the basis coefficients $a_n(t)$ can then be extracted using the orthogonality of the basis functions at a given pump power, as defined in Eq.~(\ref{orth}). Multiplying through by $\varphi_m(\mathbf{r};P,\Omega)$ and integrating over the pump region $\mathcal{P}$ leads to the following equation for $a_n(t)$:
\begin{align}
i\frac{da_n}{dt} &= \left[\nu_n(P)-\Omega\right]a_n + s\gamma_R\sum_{m}N_{nm}(t)a_m \nonumber \\
&+ g\sum_{mrs} A_{nmrs} a_ma_ra_s^*
\label{CMTCond}
\end{align}
The first line includes exclusively the effect of the reservoir. Here, the first term comes simply from the linear problem, and encapsulates the physics of reservoir-induced gain and mirror loss, as well as repulsion due to reservoir excitons. Indeed, this physics is manifest in the dependence of the eigenvalue $\nu_n(P)$ on pump power, as has been discussed earlier: a positive or negative imaginary part indicates gain or loss respectively, while the real part indicates frequency and evolves with pump power due to the repulsion-induced blueshift. 

The second term describes nonlinear condensate-reservoir interactions, which are expressed in terms of dynamical reservoir `matrix elements' $N_{nm}(t)$:
\begin{align}
N_{nm}(t) = \int_{\mathcal{P}} d\mathbf{r}~~\varphi_n~\widetilde{n}_R(\mathbf{r},t)~\varphi_m
\label{Nnm}
\end{align}
Crucially, the reservoir matrix elements are also dynamical unknowns. Thus, the set of Eqs.~(\ref{CMTCond}) cannot alone be solved for the basis coefficients; as one might expect, it is necessary to obtain the governing dynamical equation for the nonlinear, time-dependent part of the reservoir density, $\widetilde{n}_R(\mathbf{r},t)$, which can then be used to obtain an equation for $N_{nm}(t)$. This is the subject of the next section.

Before moving on, we note that the only term on the second line describes polariton-polariton interactions within the condensate, $\propto g$. The mode overlap matrix $A_{nmrs}$ modulates the strength of interactions between different modes; it has the form:
\begin{align}
A_{nmrs} = \int_{\mathcal{P}} d\mathbf{r}~\varphi_n\varphi_m\varphi_r\varphi_s^*
\end{align}
The overlap matrix elements are generally complex owing to the non-Hermitian nature of the pump modes.

\subsection{Reservoir Dynamics}

Recall that reservoir dynamics are governed by Eq.~(\ref{res}), reproduce here for clarity:
\begin{align}
\frac{\partial}{\partial t}n_R(\mathbf{r},t) = Pf(\mathbf{r}) - \gamma_R n_R - Rn_R|\Psi(\mathbf{r},t)|^2
\label{fullRes}
\end{align}
The reservoir-condensate coupling here is a density-density term, clearly of a different form compared to the reservoir-condensate coupling that appears in the gGPE, Eq.~(\ref{NGPE}). Thus, simply projecting Eq.~(\ref{fullRes}) in its present form onto the pump-dependent basis does not yield a closed set of equations for the unknowns $\{a_n(t),N_{nm}(t)\}$. Before presenting a workaround, we consider the regime of fast reservoir relaxation, where Eq.~(\ref{fullRes}) can be explicitly solved, and this problem does not arise.

\subsubsection{Fast reservoir relaxation}

When the reservoir relaxation rate $\gamma_R$ is fast compared to other timescales of condensate evolution, namely mirror-loss $\gamma_c$ [for details, see Appendix~\ref{app:resElim}], the reservoir density is `enslaved' to the condensate evolution, and as such its dynamics may be adiabatically eliminated. We can set $\dot{n}_R(\mathbf{r},t) = 0$ in Eq.~(\ref{res}), which then yields a closed form expression for the reservoir density in terms of the condensate density $|\Psi(\mathbf{r},t)|^2$:
\begin{align}
n_R^{\rm ad} = \frac{Pf(\mathbf{r})}{\gamma_R + R|\Psi(\mathbf{r},t)|^2}
\label{resNLSS}
\end{align}
If we now perform a displacement of the reservoir density as introduced in Eq.~(\ref{displaceRes}), the nonlinear, time-dependent part of the reservoir density $\widetilde{n}_R^{\rm ad}(\mathbf{r},t)$ takes the form:
\begin{align}
\widetilde{n}_R^{\rm ad}(\mathbf{r},t) = - \frac{R}{\gamma_R}\left[\frac{Pf(\mathbf{r})}{\gamma_R + R|\Psi(\mathbf{r},t)|^2} \right] |\Psi(\mathbf{r},t)|^2
\label{adiabaticRes}
\end{align}
This term represents the depletion of the static reservoir with increase in the condensate density, as evidenced by its negative-definite nature. By making use of the expansion of $\Psi(\mathbf{r},t)$ in the pump-dependent basis, Eq.~(\ref{exp}), in the above and substituting the result into Eq.~(\ref{Nnm}), the reservoir matrix elements $N_{nm}(t)$ can be expressed in terms of the basis coefficients $a_n(t)$. Therefore, in the regime of fast reservoir relaxation where Eq.~(\ref{adiabaticRes}) is valid, the full condensate dynamics are given entirely by the evolution equations for the basis coefficients, Eqs.~(\ref{CMTCond}), with $N_{nm}(t)$ also expressed in terms of these coefficients.

However, the appearance of $|\Psi|^2$ nonlinearly in the denominator of Eq.~(\ref{adiabaticRes}) means that spatial integrals need to be computed at every time step, which has significant computational overhead. Much more restrictively, adiabatic elimination is valid only when $\gamma_R \gg \gamma_c$. To allow the simulation of arbitrary dynamical regimes, we now develop an approach that is not handicapped by this restriction.

\subsubsection{Full reservoir dynamics}

To capture the full dynamics of the reservoir, we rewrite Eq.~(\ref{fullRes}) for the reservoir density so as to simplify the reservoir-condensate coupling term. To do so, we make use of the continuity equation that governs combined reservoir-condensate evolution; its derivation from Eqs.~(\ref{NGPE}) and (\ref{res}) is included in Appendix~\ref{app:continuity}, with the final equation given by:
\begin{align}
\frac{\partial}{\partial t}|\Psi|^2 + \Div{j}= Rn_R|\Psi|^2 - \gamma_c|\Psi|^2 
\label{continuity}
\end{align}
where $\vec{j}$ is the polariton current:
\begin{align}
\vec{j} = \frac{i}{2m} \left( \Psi \Grad{\Psi}^* - c.c.\right)
\label{probCurr}
\end{align}
Dependences on $(\mathbf{r},t)$ have been suppressed for clarity. The left hand side of Eq.~(\ref{continuity}) has the terms we expect from the continuity equation for a closed system, while the terms on the right are modifications due to the non-Hermitian nature of the incoherently pumped condensate: increase in the condensate density over time is due to stimulated scattering from the reservoir ($\propto R$), while population loss is primarily attributed to out-of-plane or mirror loss ($\propto \gamma_c$). The integral form of Eq.~(\ref{continuity}) in the linear regime has been shown to yield important insights into the gain optimization principle that determines the lowest threshold condensate mode~\cite{sun_stable_2016}.

More importantly for the task at hand, Eq.~(\ref{continuity}) can also be viewed as a constraint on the evolution of the product of reservoir-condensate densities. Therefore, it allows us to eliminate the product $Rn_R|\Psi|^2$ in favour of terms that depend only on the condensate density, and not $n_R$ . Upon this substitution, and the displacement transformation of Eq.~(\ref{displaceRes}), the dynamical equation for the nonlinear, time-dependent part of the reservoir density, $\widetilde{n}_R(\mathbf{r},t)$, becomes:
\begin{align}
\frac{\partial}{\partial t} \widetilde{n}_R = - \gamma_R \widetilde{n}_R - \frac{\partial}{\partial t}|\Psi|^2 -\gamma_c|\Psi|^2 - \Div{j}
\label{fullRes3}
\end{align}
The final step is to project Eq.~(\ref{fullRes3}) onto the pump basis; we again expand the condensate wavefunction $\Psi(\mathbf{r},t)$ in the pump-power dependent basis, multiply through by $\varphi_n\varphi_m$ and integrate over the pump region $\mathcal{P}$ to obtain an equation for the reservoir matrix elements $N_{nm}(t)$ (details of this derivation are included in Appendix~\ref{app:fullCMTDeriv}). The resulting equation for $N_{nm}(t)$, together with Eq.~(\ref{CMTCond}) for the mode coefficients, are:
%\begin{widetext}
\begin{subequations}
\begin{align}
i\frac{da_n}{dt} = &\left[\nu_n(P)-\Omega\right]a_n + s\gamma_R\sum_{m}N_{nm}(t)a_m \nonumber \\
&+ g\sum_{mrs} A_{nmrs} a_ma_ra_s^* \label{TCMTa} \\
\frac{dN_{nm}}{dt} =  &-\gamma_R N_{nm} - \sum_{rs} A_{nmrs}\left( i\left[ \nu_r(P) - \nu^*_s(P) \right] + \frac{d}{d t}\right) \nonumber \\
&\times (a_ra_s^*) -  \frac{R}{\gamma_R}P\sum_{rs}B_{nmrs}a_ra_s^* \label{TCMTb}
\end{align}
\end{subequations}
%\end{widetext}
Eq.~(\ref{TCMTb}) for the reservoir matrix elements may seem complicated, but the physics it represents is simply two-fold: reservoir depletion via scattering from the reservoir into the condensate, rewritten using the continuity equation, and all other reservoir loss mechanisms, encapsulated via $\gamma_R$. We have also introduced the pump overlap matrix $B_{nmrs} = \int_{\mathcal{P}}d\mathbf{r}~f(\mathbf{r})\varphi_n\varphi_m\varphi_r\varphi_s^*$. These equations constitute the TCMT, a modal description of the gGPE-reservoir equations, in our non-Hermitian pump basis. The coupled nonlinear PDEs are hence reduced to a set of nonlinear ODEs, with all the spatial information included in the overlap matrices $A_{nmrs}$ and $B_{nmrs}$. For a basis expansion featuring $N$ modes, the TCMT has $N$ equations for the basis coefficients and $N(N+1)/2$ equations for the reservoir matrix elements, where we take advantage of the fact that the reservoir matrix is symmetric under exchange of its indices, $N_{nm} = N_{mn}$ (see Eq.~(\ref{Nnm})). 

The TCMT simulation procedure is straightforward: for fixed system parameters $\{g,g_R,R,\gamma_c,\gamma_R\}$ and pump power $P$, we first compute the $N$ pump modes at that pump power, as discussed in Sec.~\ref{sec:basis}. This allows computation of the spatial overlap matrices $A_{nmrs}$ and $B_{nmrs}$ and the basis eigenvalues $\{\nu_n(P)\}$. Then, Eqs.~(\ref{TCMTa}),~(\ref{TCMTb}) can be easily simulated (once initial conditions are specified) for the basis coefficients and reservoir matrix elements. Eq.~(\ref{exp}) allows computation of the condensate wavefunction, and hence related observables, from the simulation results.

\section{Numerical Tests}
\label{sec:numerics}

We now proceed to tests of the developed modal description by comparing direct numerical simulations of the TCMT to those from a full spatio-temporal integration of the nonlinear gGPE and associated reservoir dynamics equation using a standard split-step integrator (SSI). We restrict our analysis to one dimensional geometries, where simulation times for the SSI are still feasible. However, even here, we find that the TCMT easily outperforms the SSI in simulation speed. 

For the SSI, a large spatial domain is used to avoid any numerical artefacts due to the imposition of periodic boundary conditions in the FFT-based algorithm. In contrast, the spatial modes for the TCMT are computed only within the pumped region, and only once at a given pump power. The spatial resolution is kept equal for both methods. We also employ Strang splitting for accurate time evolution with the SSI~\cite{strang_construction_1968}, while the TCMT is based on an adaptive step-size solver which is checked for stability by choosing small enough temporal step sizes. 

\subsection{1D Uniform Pump}

As a first example, we consider the uniform pump configuration that was introduced earlier (see Fig.~\ref{f:pumpRegion}~(a)), compute the \ltms{} and corresponding pump modes for the pump basis, and simulate Eqs.~(\ref{TCMTa}),~(\ref{TCMTb}) for the basis coefficients $\{a_n(t)\}$ and reservoir matrix elements $\{N_{nm}(t)\}$. Since dynamical unknowns differ between the SSI and the TCMT, performing an equivalent initialization procedure across both requires some explanation. The SSI requires choosing an initial wavefunction $\Psi(\mathbf{r},0)$, and initial reservoir density $n_{R}(\mathbf{r},0)$. To map these to equivalent initial conditions for the TCMT, $\Psi(\mathbf{r},0)$ is projected onto the basis of size $N$ being used. Then, the initial values of basis coefficients $\{ a_{n}(0) \}$ can be isolated using the orthogonality of the basis modes. For the reservoir density, we first displace the reservoir [See Eq.~(\ref{displaceRes})] to obtain $\widetilde{n}_{R}(\mathbf{r},0)$, then use Eq.~(\ref{Nnm}) to compute the initial reservoir matrix elements $\{ N_{nm}(0) \}$. Defining $\Psi \equiv |\Psi(\mathbf{r},t)|e^{-i\phi(\mathbf{r},t)}$, physically relevant quantities namely the condensate density $|\Psi(\mathbf{r},t)|^2$, phase $\phi(\mathbf{r},t)$, and the total polariton number $\rho(t) = \int d\mathbf{r}~|\Psi(\mathbf{r},t)|^2$, can be easily obtained from the TCMT results using Eq.~(\ref{exp}). 

\begin{figure}[t!]
\includegraphics[scale=0.3]{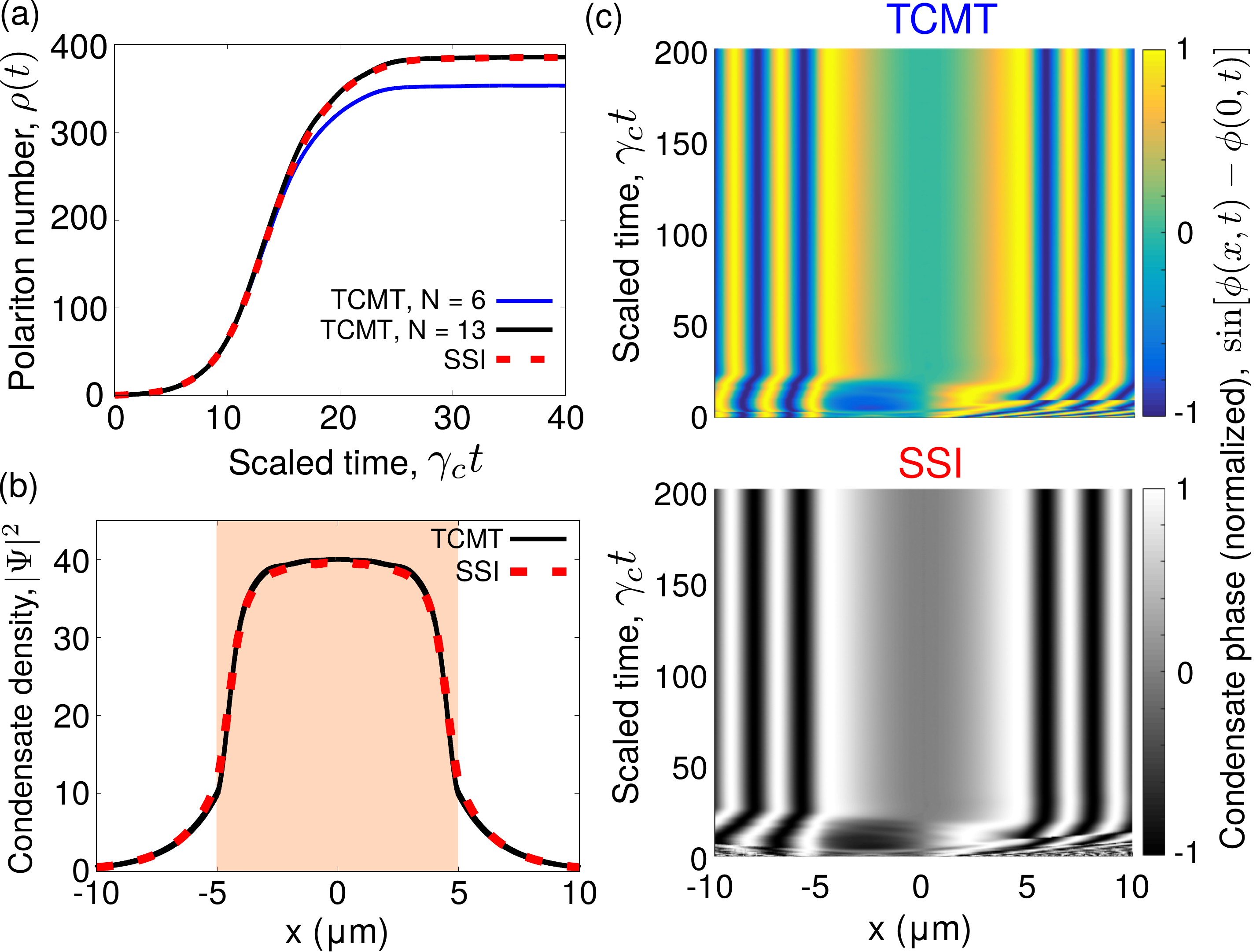}
\caption{(a) Plot of the total polariton number (integrated condensate intensity) as a function of time, under a uniform 1D pump configuration using the TCMT (solid lines) and the SSI (dashed lines). Here, the reservoir relaxation rate is fast compared to the mirror loss, $\gamma_R = 10 \gamma_c$, and the pump power $P = 1.50P_1^{\rm L}$. (b) Plot of the spatial condensate density distribution using the TCMT with $N = 13$ modes (solid black) and the SSI (dashed red). The orange shaded area indicates the pumped region. (c) Plot of (sine of the) condensate phase in space (horizontal axis) and scaled time (vertical axis), relative to the phase $\phi(x=0,t)$, for the TCMT (top panel) and SSI (bottom panel).}
\label{gammaR10Comp}
\end{figure}

We begin by considering the case of fast reservoir relaxation, where $\gamma_R$ is large compared to the mirror loss rate $\gamma_c$; in particular, $\gamma_R = 10\gamma_c = 10$~meV. The repulsive interactions are both kept on, with $g_R =~0.072$~$\mu$m\textsuperscript{2}~meV and $g = ~0.04$~$\mu$m\textsuperscript{2}~meV. The amplification rate is fixed at $R =~0.1$~$\mu$m\textsuperscript{2}~meV, and $m^{-1} =~0.59$~$\mu$m\textsuperscript{2}~meV. We follow condensate dynamics as the pump power is increased beyond the lowest linearized power threshold value, $P_1^{\rm L}$, which - as discussed earlier - is the threshold for condensation in this case. Fig.~\ref{gammaR10Comp}~(a) shows results for $P =~1.50P_1^{\rm L}$, well into the nonlinear regime, for the total polariton number as a function of time as computed using the TCMT (solid lines) and the SSI (dashed line).  Note the good agreement when a pump basis of size $N = 13$ is used. At this pump power, including additional pump modes does not change the simulation results. Encouragingly, the correct qualitative behaviour is reproduced with as few as $N = 6$ pump modes, and the inclusion of additional modes leads to improved agreement wih the SSI.  

The steady state condensate density distribution is plotted in Fig.~\ref{gammaR10Comp}~(b) for the TCMT (solid curve, $N=13$ pump modes) and the SSI (dashed curve). Again, very good agreement is seen between the two methods. The polariton distribution is significantly more delocalized than the lowest \ltm{} for this pump configuration into which condensation first occurs, shown in Fig.~\ref{f:eigenComplex}~(c). Here the modal description we have developed provides useful insight: mixing with higher order modes due to interactions leads to the modified condensate density distribution. Such a description escapes the standard gGPE. Finally, the (sine of the) condensate phase is presented in Fig.~\ref{gammaR10Comp}~(c) in both space and time, for the TMCT (top panel) and SSI (bottom panel). The phase evolution and emergent coherence agree very well across both methods. The very slight discrepancy seen in initial dynamics is not definitive: we find that here the SSI results themselves can vary when spatial grid size is changed.

\begin{figure}[t!]
\includegraphics[scale=0.62]{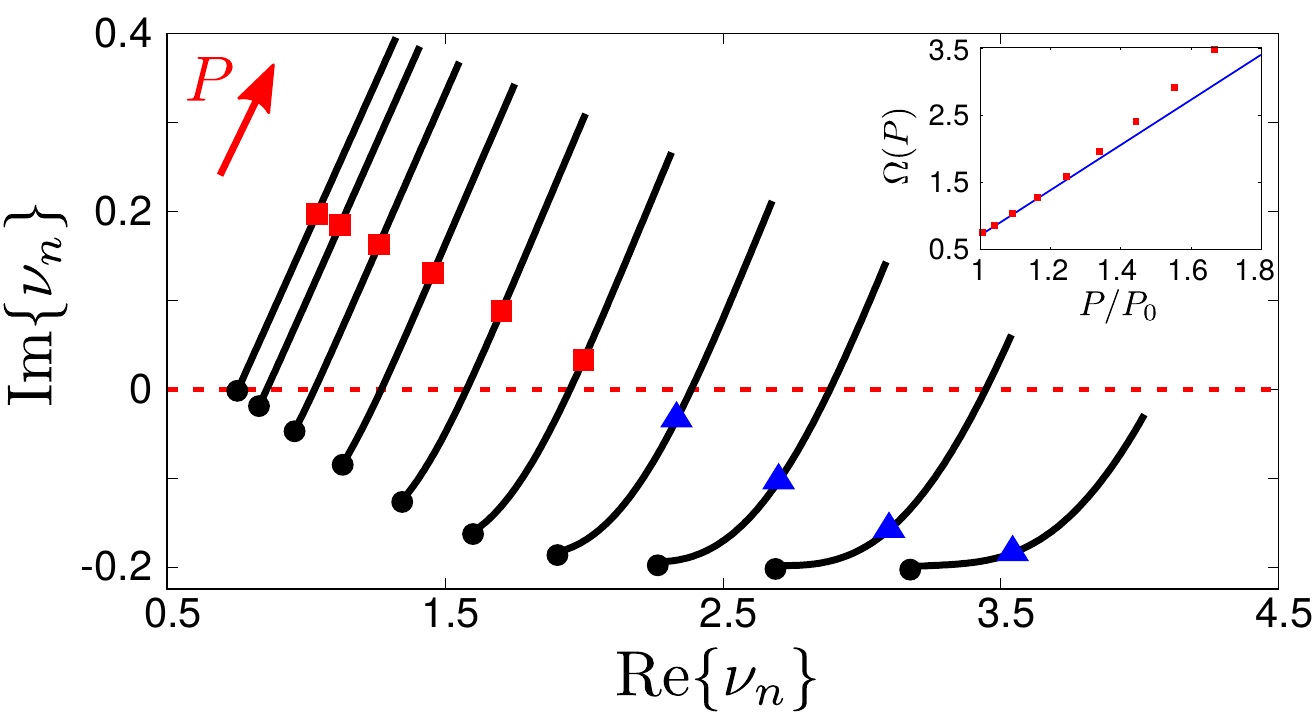}
\caption{Trajectories of non-Hermitian basis mode eigenvalues as a function of power for the case of a 1D uniform pump, with pump powers $P \in [1,2]P_1^{\rm L}$, and $\gamma_R = 10\gamma_c$. The difference from the trajectories shown in Fig.~\ref{f:eigenComplex} is that here the outgoing frequency $\Omega(P)$ is nonlinearly determined at every pump power via self-consistent solution of the TCMT; the evolution of $\Omega(P)$ is shown in the inset. Red squares indicate the frequencies $\{\omega_n^{\rm L}$\} of the \ltms{}. }
\label{TDEigenvals}
\end{figure}

The convergence of simulation results as the number of pump modes increases is a direct consequence of how these modes are defined. To clarify this, we draw attention to the plot in Fig.~\ref{TDEigenvals} of basis eigenvalue trajectories $\nu_n(P)$ as a function of pump power, for pump modes with $n \in [1,10]$. The pump-dependent outgoing frequency $\Omega(P)$ is shown in the inset; its evolution is determined by computing the frequency spectra of mode coefficients $\mathcal{F}\{a_n(t)\}$ (where $\mathcal{F}\{\cdot\}$ is the Fourier transform), which in this case yields a single-frequency condensate. As the pump power increases, the eigenvalues flow across the complex plane as before; the only difference here is the evolution of $\Omega(P)$ (shown in the inset), which modifies the flow from that shown in Fig.~\ref{f:eigenComplex}~(a), where $\Omega$ is constant. 

For the specific pump power $P = 1.50P_1^{\rm L}$ corresponding to Fig.~\ref{gammaR10Comp}, the squares and triangles indicate respectively the eigenvalues with positive or negative imaginary parts. Note that basis modes for larger $n$ have increasingly more negative imaginary parts. This is in fact explicitly due to the correspondence of the $n^{\rm th}$ pump mode with the $n$th \ltm{}: \ltms{} with larger $n$ have higher power thresholds, which is indicative of the comparatively lower gain they experience compared to \ltms{} with lower $n$. This amount of gain manifests in the imaginary parts of eigenvalues of the pump modes that correspond to these \ltms{}. Roughly speaking, this gain-like term controls the importance of the pump modes as it explicitly appears in Eq.~(\ref{TCMTa}), although since the simulated problem is no longer linear this control is not absolute. Including only pump modes with eigenvalues above the real line - the first six modes here - already shows reasonable agreement with the full solution. Generally, we find that the inclusion of pump modes corresponding to the lowest few \ltms{} is already sufficient for a qualitative understanding of dynamics; the addition of more modes simply refines the TCMT solution without incurring significant changes. As pump power increases and imaginary parts of the pump mode eigenvalues become less negative, as is clear from Fig.~\ref{TDEigenvals}, more and more pump modes need to be included in the expansion.

\begin{figure}[t!]
\includegraphics[scale=0.307]{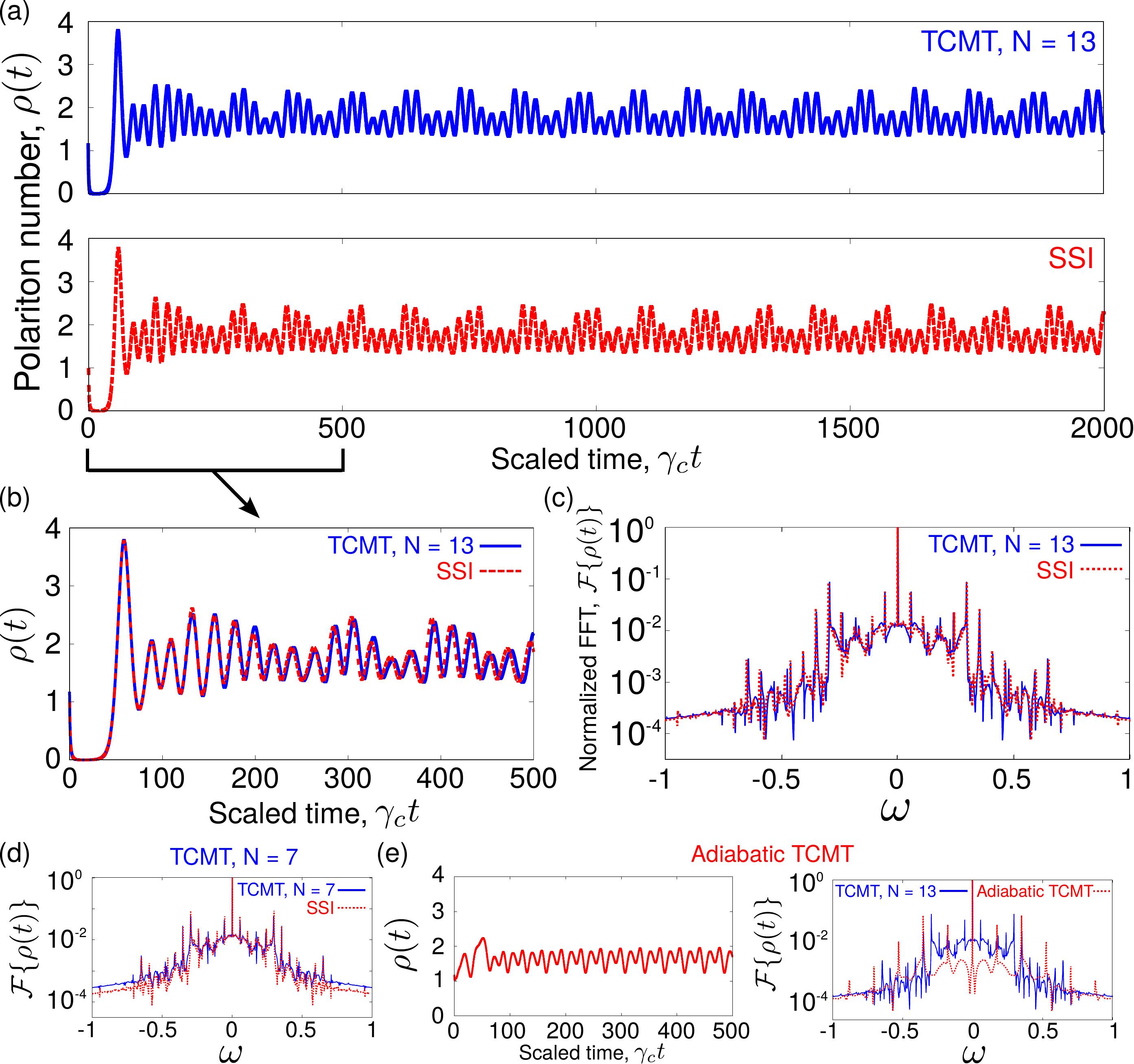}
\caption{(a) Plot of polariton number $\rho(t)$ using the TCMT with $N=13$ basis modes (top) and SSI (bottom), under a uniform pump configuration. Here, the reservoir relaxation rate is slow, $\gamma_R= 0.1 \gamma_c$, and $P = 1.30P_1^{\rm L}$. (b) Zoomed-in version of (a) for $\gamma_c t\in[0,500]$, with TCMT results in solid blue and SSI results in dashed red. (c) Plot of the frequency spectra of total polariton number $\mathcal{F}\{\rho(t)\}$, using the TCMT (solid blue), and the SSI (dashed red). (d) Plot of $\mathcal{F}\{\rho(t)\}$ using only $N = 7$ modes for the TCMT (solid blue), with the SSI (dashed red) for comparison. (e) Plot of the polariton number $\rho(t)$ and its spectrum $\mathcal{F}\{\rho(t)\}$ using the `adiabatic' TCMT (see text), where deviations from the results of (b), (c) are clearly visible.}
\label{gammaR01Comp}
\end{figure}

Next, we turn to the opposite dynamical regime, where reservoir relaxation is slow compared to the mirror loss; in particular, we take $\gamma_R = 0.1\gamma_c$. In this regime, dynamical instabilities and non-steady-state behavior is known to be prevalent~\cite{bobrovska_stability_2014}, as manifest in the nonstationary polariton number $\rho(t)$ plotted in Fig.~\ref{gammaR01Comp}~(a) for the TCMT (top panel) and SSI (bottom panel) at pump power $P = 1.30P_1^{\rm L}$. The initial dynamics are well captured, as is clear from the zoomed-in plot for $\gamma_c t\in[0,500]$ in Fig.~\ref{gammaR01Comp}~(b). Comparisons of long time dynamics are simplest in the frequency domain; we plot the normalized frequency spectra of polariton number, $\mathcal{F}\{\rho(t)\}$ in logscale in Fig.~\ref{gammaR01Comp}~(c) for the TCMT (blue) and the SSI (red). The spectra agree very well, up to slight discrepancies. For completeness, plots comparing condensate density and phase evolution in space and time for both methods are provided in Appendix~\ref{app:times}, Fig.~\ref{appGammaR01}.

When only $N=7$ pump modes are used, the resulting normalized frequency spectrum $\mathcal{F}\{\rho(t)\}$ is shown in Fig.~\ref{gammaR01Comp}~(d); this already agrees reasonably with the full result. Finally, in this regime where $\gamma_R < \gamma_c$, the often-used technique of adiabatically eliminating the reservoir density evolution should no longer be valid; our approach allows us to check this approximation. Using the `adiabatic' TCMT, where reservoir matrix elements $N_{nm}(t)$ are determined by $\widetilde{n}_R^{\rm ad}$ given by Eq.~(\ref{adiabaticRes}), we compute and plot the polariton number $\rho(t)$ and its spectrum $\mathcal{F}\{\rho(t)\}$ in Fig.~\ref{gammaR01Comp}~(e). Sure enough, we see clear discrepancies between these results and the analogous Figs.~\ref{gammaR01Comp}~(b),~(c), where full reservoir dynamics are included. Therefore, in this regime the nontrivial dynamics of the reservoir are crucial in determining the correct condensate behaviour; adiabatic elimination of the reservoir density fails to capture these dynamics, whereas our full treatment of the reservoir dynamics' equation via Eq.~(\ref{TCMTb}) does. 

Additional simulation results under uniform pumping for varying polariton-polariton interaction strengths $g$ are provided in Appendix~\ref{app:times}, together with a comparison of simulation times showing the speedup in using the TCMT relative to the SSI.

\subsection{Coupled Polariton traps under Incoherent Pumping}

\begin{figure}[t!]
\includegraphics[scale=0.135]{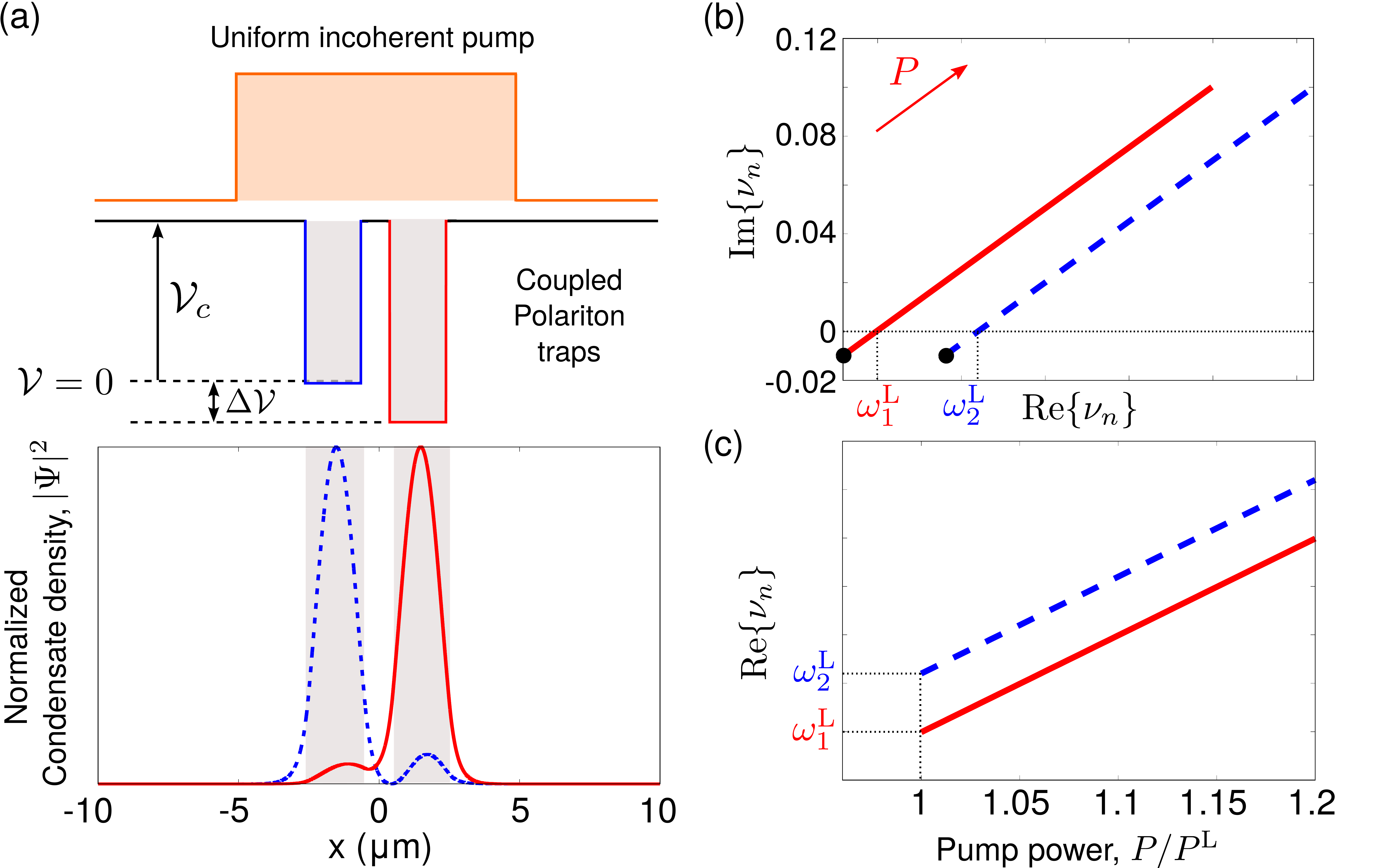}
\caption{(a) Top: geometry of the trapping potential and pump profile under consideration. $\mathcal{V}_c$ is the confining potential, and $\Delta\mathcal{V}$ is the detuning of the two traps. Bottom: Two lowest threshold modes for the geometry in the top panel. (b) Flow of eigenvalues with pump power in the complex-$\nu$ plane for the two pump modes corresponding to the \ltms{} shown in (a). (c) Evolution of the real part of basis eigenvalues as a function of pump power, showing the reservoir-induced blueshift in the linear regime.}
\label{synchGeo}
\end{figure}

The TCMT can also be used to study dynamics of incoherently pumped polaritons trapped in nontrivial potential landscapes. We will consider this situation as a second numerical example, looking at the case of polariton condensation in coupled polariton traps that can be generated by fabricating a trapping potential for their photonic component~\cite{kaitouni_engineering_2006, lai_coherent_2007, utsunomiya_observation_2008, kim_gaas_2008, kim_dynamical_2011}. Such traps can be generated by unintentional disorder as well \cite{baas_synchronized_2008}. The system is pumped incoherently with a wide beam spot; the trapping potential and pump geometry are shown in Fig.~\ref{synchGeo}~(a). For concreteness, we take a beam spot of $10~\mu$m (in comparison to trap widths of $2~\mu$m), with the two traps being separated by $1~\mu$m. The strength of the trapping potential is $\mathcal{V}_c$. We consider the case where the two traps have slightly different depths, with the difference in depth being $\Delta\mathcal{V}$. We emphasize that arbitrary potential landscapes and pump geometries may be simulated.

For a given spatial configuration of the system, the \ltms{} are computed as discussed in Sec.~\ref{sec:LTM}, but now with the addition of the trapping potential $\mathcal{V}_{\rm trap}(\mathbf{r})$ in Eq.~(\ref{linGPE}) to account for the confining geometry. Then, the new linear non-Hermitian generator for this system is defined via:
\begin{align}
\mathcal{H}_{\rm L}(P)\Psi \equiv \left[ -\frac{\nabla^2}{2m} + \mathcal{V}_{\rm trap}(\mathbf{r}) + s Pf(\mathbf{r}) - \frac{i}{2}\gamma_c \right]\Psi
\label{wellGPE}
\end{align}
Here, the solution outside $\partial\mathcal{P}$ is no longer an outgoing polariton flux, but is rather confined by the trapping potential. The boundary condition is then of the form introduced in Eq.~(\ref{bc}), but is now characterized by the wavevector $q^2(\nu_n)/2m = \nu_n + i\frac{\gamma_c}{2} - \mathcal{V}_c$, where $\mathcal{V}_c$ is the strength of the confining potential beyond $\partial\mathcal{P}$ (see Fig.~\ref{synchGeo}~(a)).

The two \ltms{} with lowest power thresholds obtained for this system are shown in the lower panel in Fig.~\ref{synchGeo} (a), in solid red and dashed blue; as expected, each mode is localized mostly in one trap. However, the coupling between the traps leads to a renormalization such that each mode has some density in the neighbouring trap. The mode confined mostly in the deeper trap (right, red) has lower frequency $\omega_1^{\rm L}$ than that confined mostly in the shallower trap (left, blue) $\omega_2^{\rm L}$. The two \ltms{} have almost equal linear threshold powers $P_1^{\rm L} \approx P_2^{\rm L} \equiv P^{\rm L}$. 

We find numerically that for low powers, a two-mode TCMT provides a complete description of condensate dynamics; including additional modes in the TCMT does not change the simulation results. As such, we will focus on a two-mode pump basis from here on, with pump modes that are connected to the $n=1,2$ \ltms{}, as described earlier. The trajectories of the eigenvalues for these basis modes as a function of pump power are plotted in the $\omega$-$\gamma$ plane in Fig.~\ref{synchGeo}~(b). Again, these eigenvalues encapsulate the physics of the linear problem: the flow towards increasing frequencies is due to the pump-induced blueshift, and the increasing imaginary part signifies the increasing gain experienced by each mode when the pump power grows. Note that the homogeneous pump spot ensures equal pump-induced blueshifts for polaritons throughout the potential landscape, and as such the two modes experience identical blueshifts; this is clear from a plot of the real part of the eigenvalues with pump power in Fig.~\ref{synchGeo}~(c). 

This system features interesting dynamical regimes that can be explored via the TCMT; for demonstrative purposes we will focus on a specific example, which is the transition from a multi-frequency condensate to a single-frequency, synchronized condensate as the pump power is increased. In our simulations, both reservoir-polariton and polariton-polariton repulsive interactions are active, $g_R = 0.2$~$\mu$m\textsuperscript{2}~meV, $g = 0.0275$~$\mu$m\textsuperscript{2}~meV, and we consider the regime of fast reservoir relaxation, $\gamma_R = 10\gamma_c = 10~$meV. The amplification rate is again $R =~0.1$~$\mu$m\textsuperscript{2}~meV and $m^{-1} =~0.59$~$\mu$m\textsuperscript{2}~meV.

\begin{figure}[t!]
\includegraphics[scale=0.21]{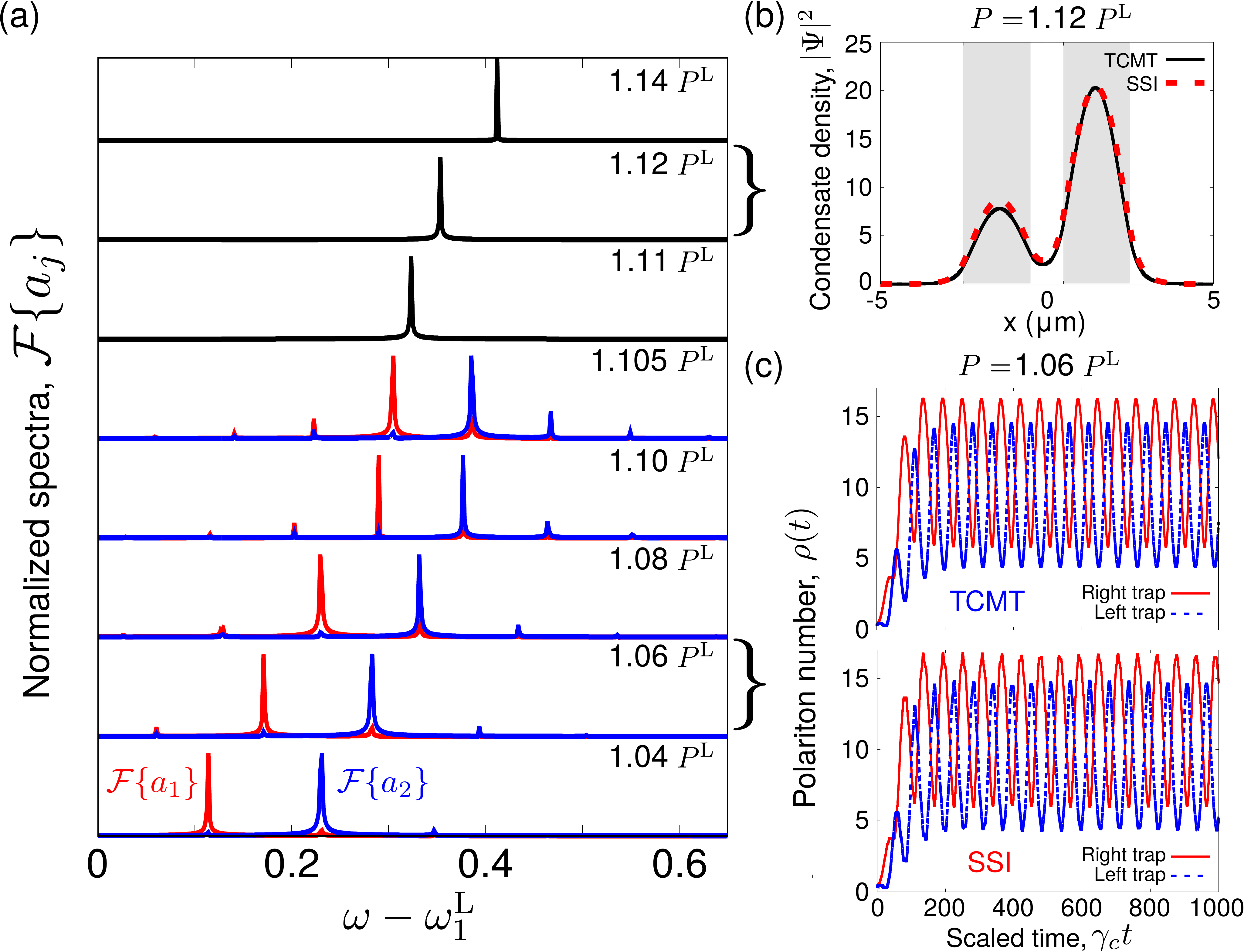}
\caption{(a) Frequency spectra of mode coefficients $\mathcal{F}\{a_j\}$ as a function of increasing pump power along the $y$-axis; each trace is labelled with the corresponding pump power. (b) Steady-state condensate density distribution as computed using the TCMT (solid black) and the SSI (dashed red), at $P = 1.12P^{\rm L}$, in the synchronized regime. The grey shaded regions indicate the locations of the polariton traps. (c) Polariton numbers in the right trap (solid red) and left trap (dashed blue) as computed using the TCMT (top panel) and the SSI (bottom panel) at $P = 1.06P^{\rm L}$, in the desynchronized regime.}
\label{synchDyn}
\end{figure}

The phenomenon of interest is clearest in frequency space; Fig.~\ref{synchDyn}~(a) shows the frequency spectra of mode coefficients $\mathcal{F}\{a_j\}$ as a function of pump power above $P^{\rm L}$, where $\mathcal{F}\{\cdot\}$ denotes the Fourier transform. Pump power increases along the $y$-axis. As pump power grows, the modes are initially distinct in frequency space, while being blueshifted due to increasing polariton-polariton repulsion. The modal frequency difference starts out being equal to the bare frequency detuning of the pump modes $\Delta\omega = \omega_2^{\rm L}-\omega_1^{\rm L}$, but is modified as pump power increases, with additional frequency components being generated. Beyond a certain threshold pump power, a transition to a synchronized regime is observed, where the modal detuning is overcome and both modes lock to a single frequency state. 

To compare these simulation results with predictions from the SSI, it is simplest to compare observables. In the multi-frequency regime, we compute the total polariton number $\rho(t)$ in each trap, which is simply the condensate density integrated over each trap's extent. Note that this is \textit{not} the same as the polariton number in each \textit{mode}, since neither basis mode is not entirely confined to just one trap. The results are shown for $P = 1.06P^{\rm L}$ in Fig.~\ref{synchDyn}~(c), where the polariton number in the right trap is plotted in solid red, and that in the left trap is shown in dashed blue. The top panel shows the TCMT result; in this regime we see that the coupling leads to an exchange of polariton densities between the two traps. The bottom panel in the same plot shows the SSI results; the agreement is quite clearly seen. As the pump power is increased and dynamics enter the synchronized regime, a time-independent steady-state is reached. For a pump power $P = 1.12P^{\rm L}$ in this regime, we plot the steady-state condensate density as obtained using the TCMT (solid black) and the SSI (dashed red) on the same plot in Fig.~\ref{synchDyn}~(b). Again, very good agreement is seen between the results from the two methods.

A number of questions particular to this system remain, for example what is the mechanism for the emergence of synchronization, and how do the strengths of various interactions affect this behaviour, to name a few. We address the rich dynamical features observed in this system in a separate publication~\cite{khan_competing_2016}. Instead, with the discussion in this section we can reasonably conclude that the TCMT is capable of accurately capturing these dynamical regimes in incoherently pumped polariton condensates, even in the presence of trapping potentials.

\begin{figure}[t!]
\includegraphics[scale=0.26]{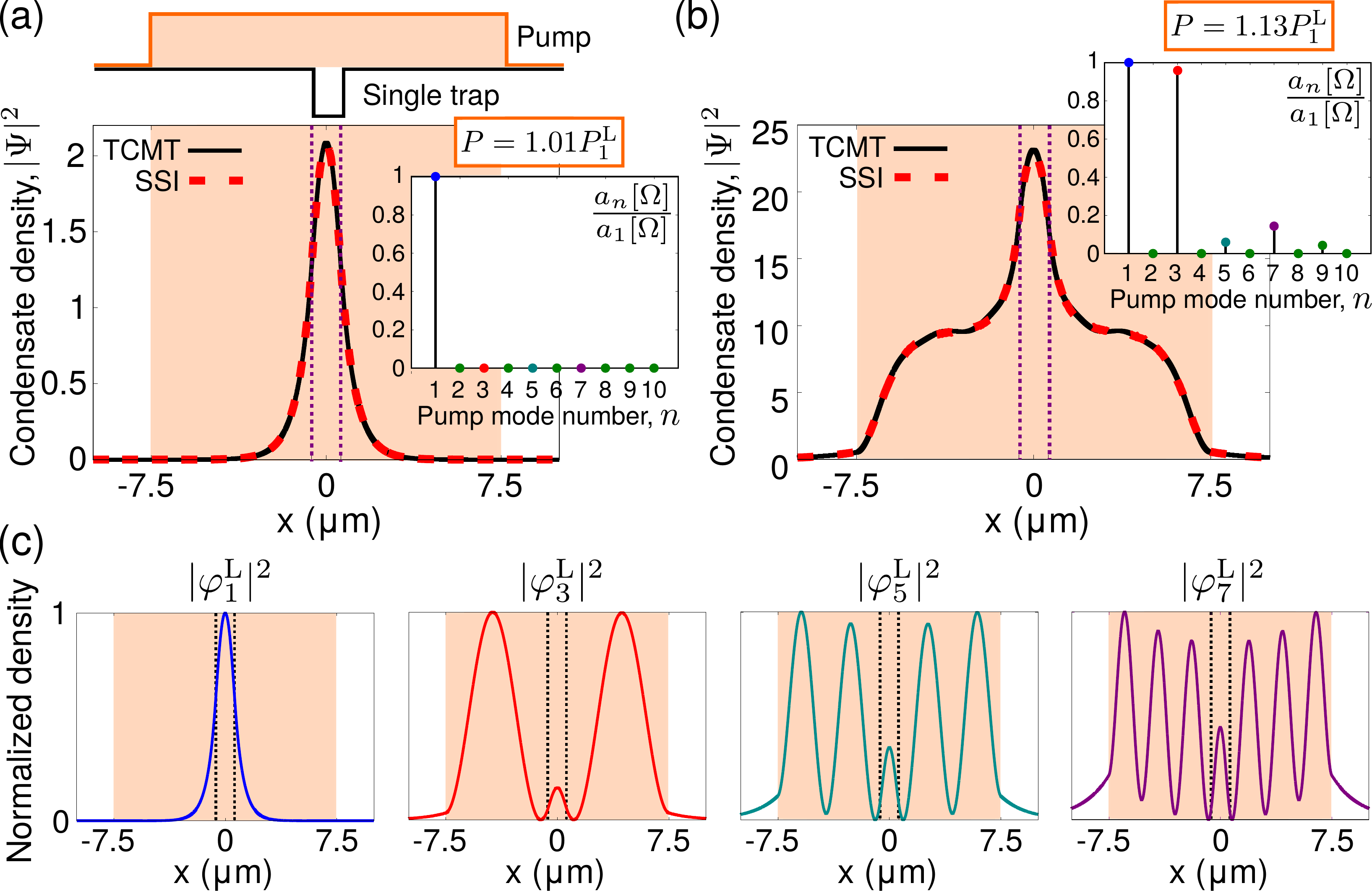}
\caption{(a) Top: System comprising a wide pump spot over a single polariton trap. Plot of the steady state condensate density $|\Psi|^2$ computed using the TCMT (solid black) and SSI (dashed red). The pump power is just above the condensation threshold, $P=1.01P_1^{\rm L}$. In this regime, $v = 0.28$ [See Eq.~\ref{v}]. (b) Steady state condensate density for $P=1.13P_1^{\rm L}$, using the TCMT (solid black) and SSI (dashed red). Here, $v \approx 1.2$, and a delocalized condensate density distribution is seen. Insets in both figures show the spectral weight of pump modes used in the TCMT. (c) Spatial density profiles of the LTMs $\{\varphi_n^{\rm L}\}$, for $n=1,3,5,7$ (from left to right). Vertical dashed lines in all figures mark the extent of the polariton trap, while orange shaded area shows the pump region.}
\label{locTransition}
\end{figure}

\subsection{Trapping vs. Nonlinear Defocusing Dynamics} 

Our previous example studied condensate dynamics in polariton traps, where the trapping potential defines the condensate modes that interact and synchronize. However, for strong enough pumping, one expects an interesting regime to appear: as the condensate density grows, the repulsive,`defocusing' potential due to nonlinear polariton-polariton interactions within the condensate ($g|\Psi|^2$) can be of the order of the linear trapping potential. In this regime, the localizing nature of the trap competes with this defocusing potential. We conclude our numerical tests of the TCMT by analyzing this regime. 

Consider the system configuration shown in the top panel of Fig.~\ref{locTransition}~(a), where a wide pump spot is incident over a single polariton trap. We employ a TCMT with $N=10$ pump modes; for a selection of these pump modes ($n=1,3,5,7$), the corresponding \ltms{} $\{\varphi_n^{\rm L}\}$ are shown in Fig.~\ref{locTransition}~(c). Note that the \ltm{} with lowest threshold, $\varphi_1^{\rm L}$, is strongly confined by the trap. However, \ltms{} for $n>1$ are more delocalized, with a majority of the mode density being outside the trap.

With growing pump power, the effect of the nonlinear potential is expected to increase; it is therefore useful to quantify the importance of the nonlinear potential \textit{relative} to the trapping potential. To do so, we introduce the quantity:
\begin{align}
v = \frac{\langle g|\Psi|^2 \rangle}{\langle\mathcal{V}(\mathbf{r})\rangle}~,~\langle G(\mathbf{r}) \rangle = \int d\mathbf{r}~G(\mathbf{r})|\Psi|^2
\label{v}
\end{align}
We use parameters employed for Fig.~\ref{gammaR10Comp}, except with a stronger polariton-polariton interaction strength, $g = 0.05~\mu$m\textsuperscript{2} meV. For a pump power just above threshold, $P=1.01P_1^{\rm L}$, a single frequency $(\Omega)$ condensate is obtained in the long time limit, with a density profile shown in Fig.~\ref{locTransition}~(a). Here, we find $v = 0.28$ for the steady state condensate; this suggests a strong effect of the trapping potential in defining the condensate profile, which is clearly seen. The TCMT's decomposition into pump modes offers further insight: the spectral weight of the coefficients $\{a_n[\Omega]\}$ corresponding to these modes is shown in the inset, normalized by $a_1[\Omega]$. Note that the only pump mode with nonzero spectral weight is that corresponding to $\varphi_1^{\rm L}$, the trapped mode.  

However, as the pump power is increased and the condensate density grows, so does the potential due to nonlinear interactions. For $P=1.13P^{\rm L}$, the condensate wavefunction is plotted in Fig.~\ref{locTransition}~(b); here, $v = 1.24$, and the effects of the defocusing potential are clear. The plot of spectral weights (inset) indicates that now the more delocalized $n>1$ pump modes also acquire weight in determining the condensate distribution observed. Note the excellent agreement with the SSI in both cases: this highlights the ability of the TCMT to capture condensate profiles even in the strongly nonlinear regime, attained dynamically with increasing pump power, using the same set of pump modes.

\subsection{Significance of the TCMT}

The excellent agreement of TCMT results with those from a split-step integration of the gGPE in different dynamical regimes provides strong evidence for the general validity of the modal description we have derived in this paper. As was seen for dynamics under a uniform pump spot, regimes with widely varying condensate dynamics (Fig.~\ref{gammaR10Comp} vs. Fig.~\ref{gammaR01Comp}) can be described using the same set of pump modes, highlighting the power of this modal representation. The judicious choice of basis is key: the correspondence of pump modes to \ltms{} lends a hierarchy of importance to the former, which manifests via the pump mode eigenvalues to control the dynamical `gain' experienced by these pump modes. An expansion in the pump basis then corresponds physically to an expansion in optimal gain-experiencing modes for a given pumping configuration, and is well-controlled with respect to the inclusion of additional modes. The TCMT can also be employed to great effect in the presence of trapping potentials, and interesting dynamical transitions can be captured. Thus, the modal theory has promise in providing both numerical and analytic insight into the physics of incoherently pumped polariton condensates, beyond what is accessible directly via ithe numerical integration of the gGPE. 

The other major advantage of the TCMT is the faster simulation speed. A comparison of simulation times is included in Appendix~\ref{app:times}; generally, the TCMT is between 1-2 orders of magnitude faster than the SSI in simulating the gGPE, in even the simplest geometries. By requiring spatial modes to be computed only once at a given pump power, the TCMT gains a significant computational advantage over the SSI, wherein the spatial wavefunction is computed at every time step. For higher dimensions, the disparity only grows: SSI computation time for a $d$-dimensional geometry grows exponentially with $d$ - specifically, $N_g^d \log_2(N_g)$, where $N_g$ is the number of grid points in every spatial coordinate (assumed equal)~\cite{bader_solving_2013}. The TCMT does not suffer from this scaling. Furthermore, the TCMT is better equipped for the computation of any quantities that require spatial integration, for example the total polariton number $\rho(t)$. While the SSI requires a spatial integral at every time step for these calculations, the TCMT's expansion in spatial modes reduces spatial integration to a simple matrix multiplication with time-independent matrix elements, which can be significantly faster~(see Appendix~\ref{app:times}).

The ability to formulate the TCMT for arbitrary pumping profiles and trapping geometries in one- and two-dimensions, together with the computational advantages the modal description provides, make the TCMT very versatile. For example, the study of dynamics in two-dimensional condensates is hindered by the computational expense of spatio-temporal integration of the gGPE. In Ref.~\cite{sun_stable_2016}, we employed the TCMT to efficiently analyze pump-dependent dynamics of polaritons confined entirely by a two-dimensional annular pump. In addition, our study of condensation in coupled traps here and further in Ref.~\cite{khan_competing_2016} lays the groundwork for application of the TCMT to condensates in lattice geometries that have garnered much interest recently~\cite{baboux_bosonic_2016}. Overall, our exploration of spatio-temporal dynamics in varying setups provides a firm foundation for the TCMT to be applied to a diverse range of polariton systems.

\section{Conclusions}

In this paper, we have introduced a modal description of polariton condensation under incoherent pumping, based on modes of the generator of linear dynamics. These non-Hermitian modes incorporate the reservoir-induced repulsive potential, and the pumping and losses inherent to polariton condensates. We have also developed a pump-dependent basis of non-Hermitian pump modes that allows an expansion of the condensate wavefunction with time-dependent basis coefficients. A temporal coupled-mode theory using this expansion provides dynamical equations for the basis coefficients and captures condensates and reservoir dynamics in the presence of all nonlinear interactions. The efficiency of our formulation was shown by successful comparisons with a full split-step integration of the generalized Gross-Pitaevskii equation, for pump-confined polariton condensates and condensation in trapping potentials. The diverse dynamical regimes we explore and the resulting agreement place the modal description on solid ground for application to other polariton condensate systems, in particular polariton lattices and two-dimensional condensates where split-step integration of the generalized Gross-Pitaevskii equation is unwieldy and less insightful.

\section{Acknowledgements}

This work was supported by the US Department of Energy, Office of Basic Energy Sciences, Division of Materials Sciences and Engineering under Grant No. DE-SC0016011.

\begin{center}
\rule{40mm}{1pt}
\end{center}

\appendix

\section{Linearized GPE}
\label{app:linGPE}

To simplify the equations describing condensate dynamics below the threshold of condensation, we will essentially linearize the dynamics in the condensate density $|\Psi|^2$, which is vanishing below the threshold power of condensation, representing the uncondensed state. Under this linearization, the equation describing reservoir dynamics [cf. Eq.~(\ref{res})] can be simplified to:
\begin{align}
\frac{\partial}{\partial t}n_R(\mathbf{r},t) = Pf(\mathbf{r}) - \gamma_R n_R
\end{align}
Linearization of this equation in $|\Psi|^2$ amounts to neglecting the effects of reservoir depletion before condensation has occurred, as must be the case. We now further consider the case where below the condensation threshold, the reservoir density has reached a steady state. In this case, the time derivative in the above equation is neglected, and we find the linear, time-independent part of the reservoir density, $n_R^{\rm L}(\mathbf{r})$:
\begin{align}
n_R^{\rm L}(\mathbf{r}) = \frac{Pf(\mathbf{r})}{\gamma_R}
\label{linRes}
\end{align}
This is precisely the part of the reservoir density that we extract explicitly in Eq.~(\ref{displaceRes}). We can similarly linearize the full, nonlinear GPE, Eq.~(\ref{NGPE}):
\begin{align}
i\frac{\partial}{\partial t}\Psi(\mathbf{r},t) = \left[ -\frac{\nabla^2}{2m} + g_Rn_R^{\rm L}(\mathbf{r}) + \frac{i}{2} \left(Rn_R^{\rm L}(\mathbf{r}) - \gamma_c\right) \right]\Psi
\end{align}
Note that we have dropped the nonlinear polariton-polariton interaction $\propto g$, and replaced the reservoir density by its linear, time-independent (undepleted) form. Using the explicit form of $n_R^{\rm L}(\mathbf{r})$ from Eq.~(\ref{linRes}) in the above leads to Eq.~(\ref{linGPE}).

\section{Practical determination of \ltms{} and basis modes}
\label{app:compModes}

Our approach to determining the \ltms{} focuses on finding modes with eigenvalues very close to the real axis, since these will be the threshold modes. We consider finely spaced windows of outgoing frequency $\Omega$, and solve for a small number of eigenvalues $\nu_n$ at every pump power for each of these windows. Only eigenvalues for which the real part, $\omega_n$ is closer to the outgoing frequency for their window $\Omega_n$, and that are close to the real line are kept; the others are discarded. Then, the pump power is increased, and the procedure is repeated. The newly computed eigenvalues are compared against the previous eigenvalues to see if these eigenvalues have crossed the real line, by checking the change in sign of their imaginary part. By choosing a small enough pump increment and fine outgoing frequency windows, the eigenmodes that cross the real line after a small power increment can be isolated accurately.

Once the \ltms{} are computed for a given system configuration, obtaining the pump modes corresponding to those \ltms{} is a relatively straightforward task. First, we note that the $P$ and $\Omega$ dependence of the pump modes themselves (\textit{not} their eigenvalues) is relatively weak, certainly for the cases we consider in this paper. Thus, even for values of $P$ and $\Omega$ different from the linear threshold values $P_n^{\rm L}$ and $\omega_n^{\rm L}$ respectively, the $n^{\rm th}$ pump mode retains its relationship to the $n^{\rm th}$ \ltm{}. This relationship is determined quantitatively by the overlap integral defined as:
\begin{align}
O_n(P,\Omega) = \int_{\mathcal{P}} d\mathbf{r}~\varphi_n^{\rm L}(\mathbf{r};P_n,\omega_n)\varphi_n(\mathbf{r};P,\Omega)
\end{align}
When $P = P_n^{\rm L}$, $\Omega = \omega_n^{\rm L}$, the $n^{\rm th}$ pump mode is identical to the $n^{\rm th}$ \ltm{}, and therefore $O_n = 1$. For all other values of $(P,\Omega)$, the $n^{\rm th}$ basis mode minimizes $|O_n(P,\Omega) - 1|$. Therefore, once the $n^{\rm th}$ \ltm{} is computed, computing the corresponding pump mode is the simple matter of finding the mode that minimizes the overlap integral $O_n$ for that \ltm{}, at a given $(P,\Omega)$ pair.

\section{Continuity equation}
\label{app:continuity}

For the system of incoherently pumped exciton-polaritons, we can derive a continuity equation describing the evolution of the condensate density as a result of the gain and loss experienced by the condensate. In particular, we look to describe the dynamics of $|\Psi|^2$, where $\Psi = \Psi(\mathbf{r},t)$ in what follows. Recall that:
\begin{align}
\frac{\partial}{\partial t}|\Psi|^2 = \Psi^*\frac{\partial}{\partial t}\Psi + c.c.
\end{align}
The time evolution of $\Psi$ is given by the generalized Gross-Pitaevskii equation, which allows us to expand the above:
\begin{align}
\frac{\partial}{\partial t}|\Psi|^2 = &-i\Bigg\{ \left[ -\frac{\Psi^*\nabla^2\Psi}{2m}\right] +\left[ \frac{i}{2}\left(Rn_R - \gamma_c\right)\right]|\Psi|^2 \nonumber \\
&+\left[g_Rn_R + g|\Psi|^2\right]|\Psi|^2  \Bigg\} + c.c.
\end{align}
We observe that the second line has a purely imaginary contribution (due to the imaginary unit up front). The right hand side of the above equation should be purely real, being the time-derivative of a real quantity, and so we are left with the simple result:
\begin{align}
\frac{\partial}{\partial t}|\Psi|^2 = -\frac{i}{2m}\left[ -\Psi^*\nabla^2\Psi + \Psi\nabla^2\Psi^* \right] + \left(Rn_R - \gamma_c\right)|\Psi|^2 
\end{align}
If we now define the polariton current $\vec{j}$ as in Eq.~(\ref{probCurr}) in the main text, the above may be rewritten as:
\begin{align}
\frac{\partial}{\partial t}|\Psi|^2 = -\vec{\nabla} \cdot \vec{j} + Rn_R|\Psi|^2 -\gamma_c|\Psi|^2
\end{align}
which can be rearranged to obtain the continuity equation as expressed in Eq.~(\ref{continuity}) of the main text.

\section{Details of derivation of coupled-mode equations}
\label{app:fullCMTDeriv}

In this appendix section, we present some additional details of derivations of the coupled-mode equations, specifically for the reservoir dynamics' equation. The first step is a simplification of Eq.~(\ref{fullRes}), for which we rearrange the continuity equation, Eq.~(\ref{continuity}):
\begin{align}
Rn_R |\Psi|^2 = \frac{\partial}{\partial t}|\Psi|^2 + \gamma_c|\Psi|^2 + \Div{j}
\end{align}
Here, the absence of the reservoir density $n_R$ on the right hand side allows us to decouple the reservoir-condensate coupling term in Eq.~(\ref{fullRes}). The effect of the reservoir is encapsulated in the evolution of the condensate wavefunction via the nonlinear GPE, and is therefore fully accounted for here. With this expression, Eq.~(\ref{fullRes}) simplifies to:
\begin{align}
\frac{\partial}{\partial t}n_R = Pf(\mathbf{r}) - \gamma_R n_R - \frac{\partial}{\partial t}|\Psi|^2 -\gamma_c|\Psi|^2 - \Div{j}
\label{fullRes2}
\end{align}
This form affords another advantage: we can now choose to displace the reservoir density as in Eq.~(\ref{displaceRes}), finally finding Eq.~(\ref{fullRes3}) of the main text. Applying this transformation to Eq.~(\ref{fullRes}) would introduce additional terms $\propto |\Psi|^2$ due to the complicated reservoir-condensate coupling term; this problem does not afflict Eq.~(\ref{fullRes2}). 

In simplifying the final form of reservoir dynamics' equation, Eq.~(\ref{fullRes3}), we first note that it is possible to rewrite the divergence of the polariton current as:
\begin{align}
\Div{j} = \frac{RPf(\mathbf{r})}{\gamma_R}|\Psi|^2 + i \left[ \Psi^* \mathcal{H}_{\rm L}(P)\Psi  - c.c. \right] - \gamma_c|\Psi|^2
\label{d3}
\end{align}
This is most easily verified by substituting the full form of $\mathcal{H}_{\rm L}(P)$ from Eq.~(\ref{linGPE}) into the above expression. We emphasise here that although written in terms of the linear generator $\mathcal{H}_{\rm L}(P)$, the condensate wavefunction above is the solution to the full nonlinear problem; we have simply rewritten $\nabla^2$ in terms of $\mathcal{H}_{\rm L}(P)$.

After substituting Eq.~(\ref{d3}) into Eq.~(\ref{fullRes3}), we have:
\begin{align}
\frac{\partial}{\partial t}\widetilde{n}_R = &-\gamma_R\widetilde{n}_R - \frac{\partial}{\partial t}|\Psi|^2 - \frac{R}{\gamma_R}Pf(\mathbf{r}) |\Psi|^2 \nonumber \\
&- i \left[ \Psi^* \mathcal{H}_{\rm L}(P)\Psi  - c.c.\right]
\label{d4}
\end{align}
Our aim is to isolate equations of motion for the reservoir matrix elements $N_{nm}(t)$; to do so, we multiply the above equation by the product of basis functions $\varphi_n\varphi_m$, and integrate over the pump region $\mathcal{P}$. First, we apply this procedure to the third and fourth terms in the first line of Eq.~(\ref{d4}) separately (the first two terms are trivial):
\begin{align}
\int_{\mathcal{P}}d\mathbf{r}~\frac{\partial}{\partial t}|\Psi|^2\varphi_n\varphi_m &= \sum_{rs} \left( \int_{\mathcal{P}} d\mathbf{r}~\varphi_n\varphi_m\varphi_r\varphi_s^* \right) \frac{d}{dt} \left(a_ra_s^*\right)\nonumber \\ 
&= \sum_{rs} A_{nmrs}\frac{d}{dt} \left(a_ra_s^*\right) \nonumber \\
\int_{\mathcal{P}} d\mathbf{r}~f(\mathbf{r})|\Psi|^2 \varphi_n\varphi_m &= \sum_{rs} \left(\int_{\mathcal{P}} d\mathbf{r}~f(\mathbf{r})\varphi_n\varphi_m\varphi_r\varphi_s^* \right)a_ra_s^* \nonumber \\
&\equiv \sum_{rs} B_{nmrs}a_ra_s^*,
\end{align}
In the last two lines, the term in brackets is used to define the pump-mode overlap matrix $B_{nmrs}$. The only remaining term in Eq.~(\ref{d4}) is that involving $\mathcal{H}_{\rm L}(P)$, in the second line. Using the expansion of $\Psi$ and the action of $\mathcal{H}_{\rm L}(P)$ on its eigenmodes as defined in Eq.~(\ref{HLP}), we have:
\begin{align}
&\int_{\mathcal{P}} d\mathbf{r}~\varphi_n\varphi_m\Psi^*\mathcal{H}_{\rm L}(P)\Psi \nonumber \\
&= \sum_{rs} \left( \int_{\mathcal{P}} d\mathbf{r}~\varphi_n\varphi_m\varphi_s^*\mathcal{H}_{\rm L}(P)\varphi_r\right) a_ra_s^* \nonumber \\
&= \sum_{rs} \left( \int_{\mathcal{P}} d\mathbf{r}~\varphi_n\varphi_m\varphi_r\varphi_s^*\right) \nu_r(P)a_ra_s^* \nonumber \\
&= \sum_{rs}A_{nmrs}\nu_r(P)a_ra_s^* 
\end{align}
The term $\Psi\mathcal{H}_{\rm L}^*(P)\Psi^*$ can be treated in the same way. Therefore, all terms from Eq.~(\ref{d4}) have been expressed in terms of basis coefficients and their spatial dependence has been integrated out. Consolidating these expressions leads to Eq.~(\ref{TCMTb}).

\section{Adiabatic elimination of reservoir density}
\label{app:resElim}

In the main text, we adiabatically eliminate the reservoir density for some calculations. For completeness, we discuss the dynamical regime where this technique is valid. For this to be the case, we require the adiabatic elimination to hold self-consistently, such that $\dot{n}_R^{\rm ad} = 0$. Hence, we consider the adiabatically eliminated value of $n_R^{\rm ad}$ (see Eq.~(\ref{resNLSS} of the main text) and compute its time derivative:
\begin{align}
\frac{\partial}{\partial t} n_R^{\rm ad} &= \frac{Pf(\mathbf{r})}{\gamma_R} \left[1 + \frac{R}{\gamma_R} |\Psi|^2 \right]^{-2} \left\{ -\frac{R}{\gamma_R} \frac{\partial}{\partial t}|\Psi|^2 \right\}  \nonumber \\
&= \left(\frac{\gamma_c}{\gamma_R}\right) \left[-\frac{Pf(\mathbf{r})}{\gamma_c\gamma_R/R}\right] \left[1 + \frac{R}{\gamma_R} |\Psi|^2 \right]^{-2} \frac{\partial}{\partial t}|\Psi|^2
\end{align}
If all other parameters are kept constant, then having $\gamma_c/\gamma_R \ll 1$ suppresses the right hand side and therefore the time derivative of $n_R^{\rm ad}$ is small. This is the fast reservoir relaxation regime where the reservoir density may be adiabatically eliminated. Here, the reservoir density is \textit{not} time-independent, rather its time-dependence is determined entirely by that of the condensate density $|\Psi(\mathbf{r},t)|^2$.

\begin{figure}[h]
\includegraphics[scale=0.28]{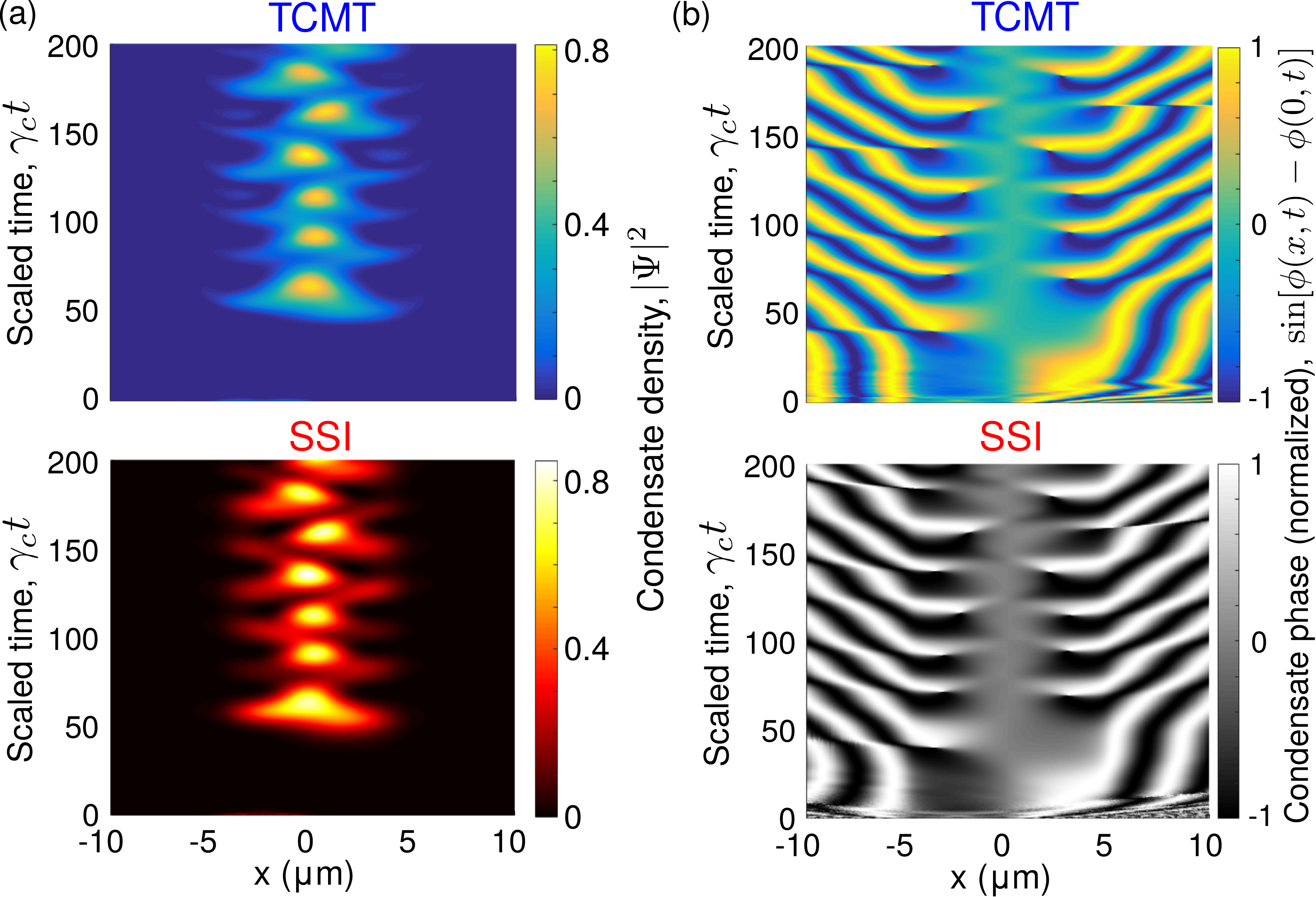} 
\caption{(a) Plot of the condensate intensity $|\Psi|^2$ as a function of space ($x$-axis) time ($y$-axis), under a uniform 1D pump configuration using the TCMT (top panel), and the SSI (bottom panel), corresponding to Fig.~\ref{gammaR01Comp}. The reservoir relaxation rate is $\gamma_R= 0.1 \gamma_c$, and $P = 1.30P_1^{\rm L}$. (b) Plot of the normalized phase using the TCMT (top) and the SSI (bottom).}
\label{appGammaR01}
\end{figure}

\section{Additional simulation results}
\label{app:times}

\begin{figure}[t!]
\includegraphics[scale=0.3]{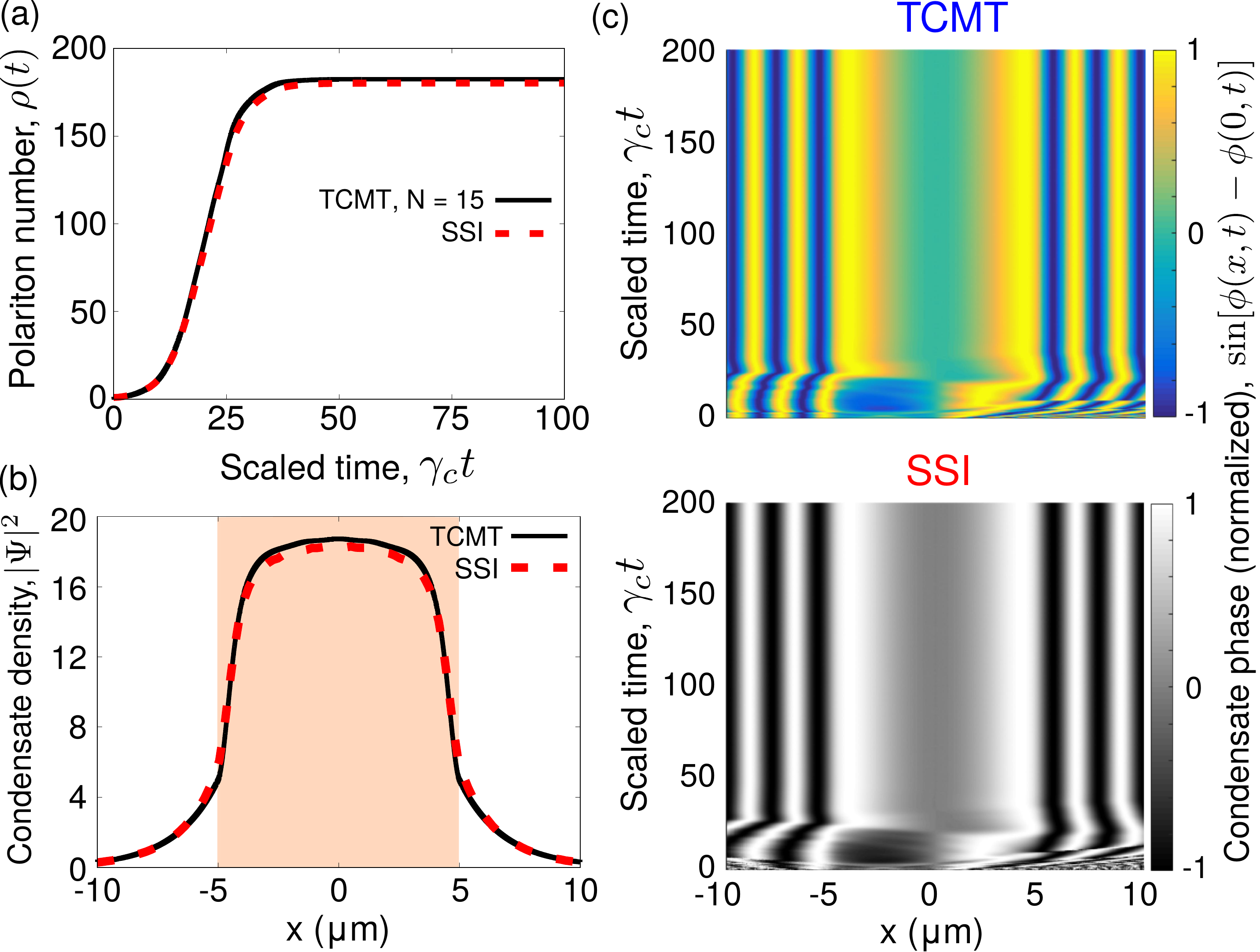}
\caption{(a) Plot of $\rho(t)$ as a function of time, under a uniform 1D pump configuration using the TCMT (solid lines) and the SSI (dashed lines). Here, $\gamma_R = 10 \gamma_c$, $P = 1.30P_1^{\rm L}$, $g_R = 0.072~\mu {\rm m}^2$~meV, and $g = 0.1~\mu {\rm m}^2$~meV. (b) Plot of steady-state $|\Psi|^2$ using the TCMT with $N = 15$ modes (solid black) and the SSI (dashed red). The orange shaded area indicates the pumped region. (c) Plot of (sine of the) condensate phase in space (horizontal axis) and scaled time (vertical axis), relative to the phase $\phi(x=0,t)$, for the TCMT (top panel) and SSI (bottom panel).}
\label{appG01}
\end{figure}

\begin{figure}[t!]
\includegraphics[scale=0.3]{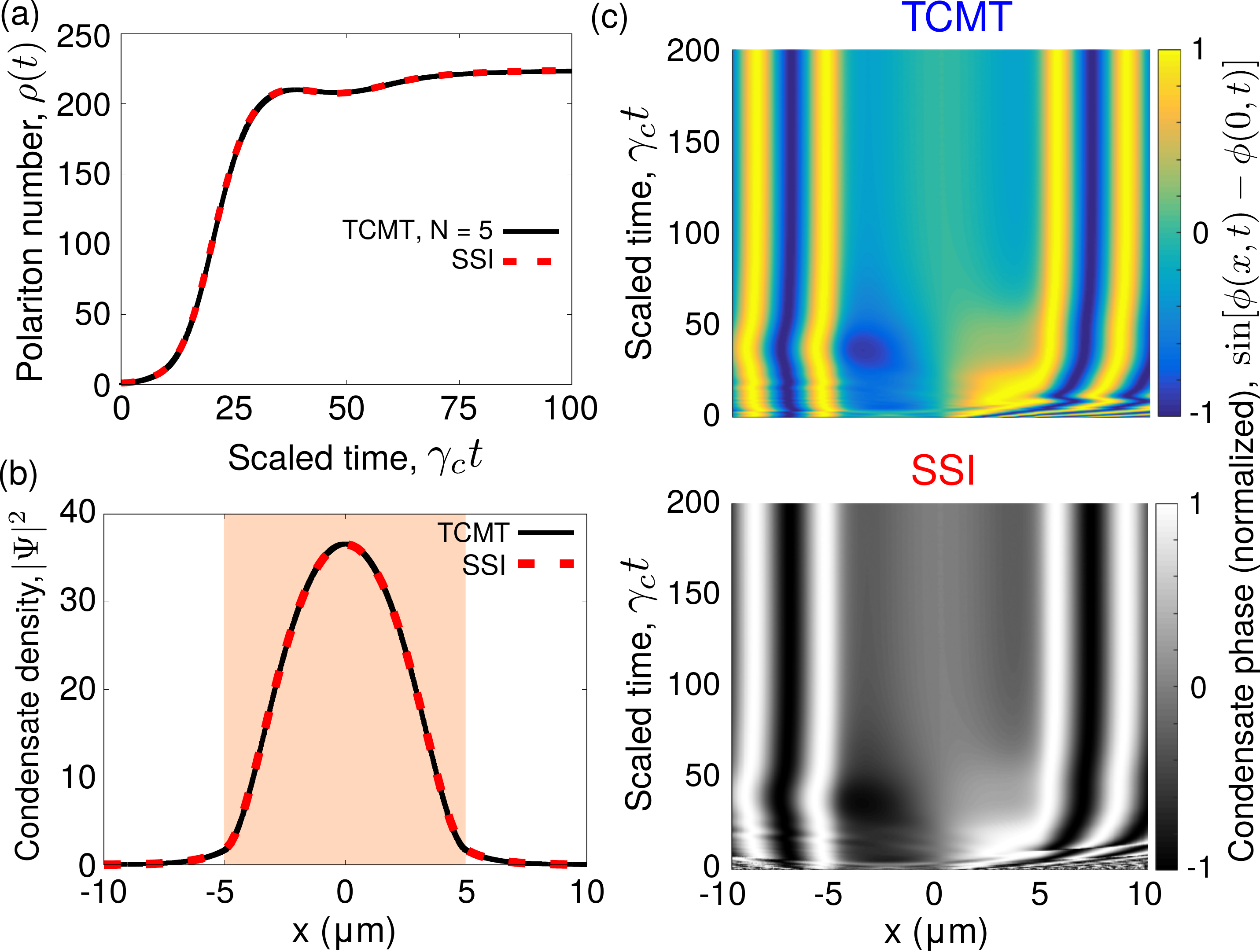}
\caption{(a) Plot of $\rho(t)$ as a function of time, under a uniform 1D pump configuration using the TCMT (solid lines) and the SSI (dashed lines). Here, $\gamma_R = 10 \gamma_c$, $P = 1.30P_1^{\rm L}$, $g_R = 0.072~\mu {\rm m}^2$~meV, and $g = 0.0072~\mu {\rm m}^2$~meV. (b) Plot of steady-state $|\Psi|^2$ using the TCMT with $N = 5$ modes (solid black) and the SSI (dashed red). The orange shaded area indicates the pumped region. (c) Plot of (sine of the) condensate phase in space (horizontal axis) and scaled time (vertical axis), relative to the phase $\phi(x=0,t)$, for the TCMT (top panel) and SSI (bottom panel).}
\label{appG00072}
\end{figure}

This appendix section includes some additional numerical results to supplement discussions from the main text. First, we provide additional simulation results in the regime of slow reservoir relaxation studied in Fig.~\ref{gammaR01Comp} of the main text, where $\gamma_R = 0.1\gamma_c$. Fig.~\ref{appGammaR01}~(a) compares the condensate density as a function of space and time for the TCMT against the SSI, for the 1D uniform pump configuration. All other parameters are the same as in Fig.~\ref{gammaR01Comp}. The result is shown for $N=13$ pump modes in the TCMT. Fig.~\ref{appGammaR01}~(b) includes plots of the (sine of the) condensate phase in space and time. We see that the complicated time dynamics in this regime are well captured by the TCMT.

Next, we provide results comparing the strongly nonlinear regime, $g = 0.1~\mu{\rm m}^2$~meV relative to $g_R = 0.072~\mu{\rm m}^2$~meV in Fig.~\ref{appG01}, to the weakly nonlinear regime, $g = 0.0072~\mu{\rm m}^2$~meV, in Fig.~\ref{appG00072}. We plot the the polariton number $\rho(t)$, the steady state condensate density $|\Psi|^2$, and (sine of the) condensate phase $\phi(x,t)$, all in the fast reservoir relaxation regime, $\gamma_R=10\gamma_c$. The excellent agreement between TCMT and SSI is clear. Also, as expected, for more strongly nonlinear regimes, the final wavefunction can be more strongly modified in comparison to the lowest threshold mode; this typically requires the incorporation of more modes in the TCMT basis. For the strongly nonlinear regime, we use $N=15$ modes, while for the weakly nonlinear regime, as few as $N=5$ modes suffice.

\begin{figure}[t!]
\includegraphics[scale=0.442]{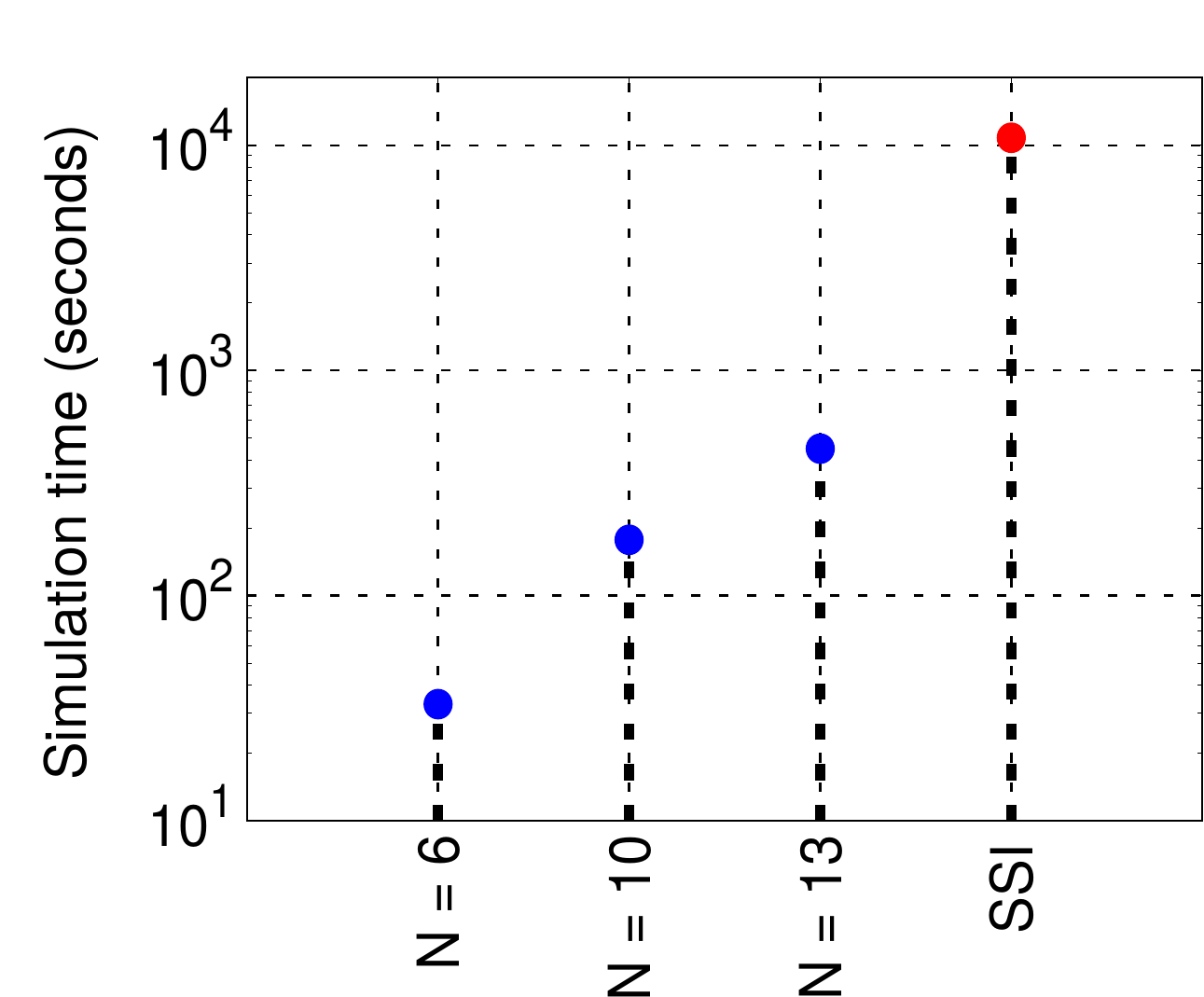}
\caption{Simulation time comparison for the simulations included in Fig.~\ref{gammaR10Comp}, for the TCMT with mode numbers $N \in $~\{6,10,13\}, and the SSI. For $N = 6$, the TCMT is more than two orders of magnitude faster than the SSI.}
\label{simTimes}
\end{figure}

To conclude, we present a comparison of simulation times for the TCMT and the SSI in Fig.~\ref{simTimes}, for the 1D uniform pump case presented in the main text, in the regime of fast reservoir relaxation. The TCMT simulation times grow with the number of pump modes $N$ being included. However, we find that for the simulations we have considered, the TCMT is still at least an order of magnitude faster than the SSI, even for the very simple 1D uniform pump geometry. We find that when more complicated pump geometries are simulated, or in the presence of nontrivial trapping potentials, the SSI can be much more inefficient at computing dynamics, whereas a few-mode TCMT may be much faster and significantly more useful.

Finally, we explain why the TCMT is much faster for the computation of spatially integrated quantities, such as the total polariton number $\rho(t)$. Consider an expansion of $\Psi$ in a pump basis of size $N$. Then, $\rho(t)$ in a region $\mathcal{R}$ may be written as:
\begin{align}
\rho(t) &= \int_{\mathcal{R}} d\mathbf{r}~|\Psi|^2 = \sum_{nm} \int_{\mathcal{R}} d\mathbf{r}~a_na_m^*\varphi_n\varphi_m^* \nonumber \\
        &= \sum_{nm} a_n \underbrace{\left[\int_{\mathcal{R}} d\mathbf{r}~\varphi_n\varphi_m^* \right]}_{C_{nm}} a_m^*
\end{align} 
As written, the above is simply a prescription for matrix multiplication; we may write:
\begin{align}
\rho(t) = 
\begin{pmatrix}
a_1 & a_2 & \ldots & a_N 
\end{pmatrix}
\begin{pmatrix}
C_{11} & C_{12} & \ldots & C_{1N} \\
C_{21} & C_{22} & \ldots & C_{2N} \\
\vdots \\
C_{N1} & C_{N2} & \ldots & C_{NN}
\end{pmatrix}
\begin{pmatrix}
a_1^* \\
a_2^* \\
\vdots \\
a_N^* 
\end{pmatrix}
\end{align}
The matrix elements $C_{nm}$ are all time-independent complex numbers which are computed only once at a given pump power.

%\nocite{*}

%\bibliography{Polariton_TCMT.bib}

%\section*{Acknowledgements}
%
%We thank XXX for stimulating discussions. This work is supported by YYY.
%
%\section*{Author Contributions}
% ZZZ
%
%\section*{Additional Information}
%Supplementary information is available in the online version of the paper. Reprints and permissions information is available online at www.nature.com/reprints. 
%
%\section*{Competing financial interests}
%The authors declare no competing financial interests.

\end{document}